\documentclass{jfm}
\usepackage{graphicx}
\usepackage{amsmath,amssymb}

\usepackage{color}
\shorttitle{Turbulent 2.5 dimensional dynamos}
\shortauthor{K. Seshasayanan and A. Alexakis}

\title{Turbulent 2.5 dimensional dynamos}

\author{K. Seshasayanan\aff{1}\corresp{\email{skannabiran@lps.ens.fr}} and A. Alexakis\aff{1}}

\affiliation{\aff{1}Laboratoire de Physique Statistique, {\'E}cole Normale Sup{\'e}rieure, CNRS UMR 8550, Universit{\'e} Paris Diderot, Universit{\'e} Pierre et Marie Curie, 24 rue Lhomond, 75005 Paris, France}

\begin{document}

\maketitle

\begin{abstract}
We study the dynamo instability driven by a turbulent two dimensional flow with three components
of the form $\left( u(x, y, t), v(x, y, t), w(x, y, t) \right)$ sometimes referred to as a 2.5 dimensional flow. 
This type of flows provides an approximation to very fast rotating flows often observed in nature.
The low dimensionality of the system allows the investigation of a wide range of fluid Reynolds number $Re$, magnetic Reynolds number $Rm$ and forcing length scales relative to the domain size 
that is still prohibited for full three dimensional numerical simulations. 
We were thus able to determine the properties of the dynamo onset as a function of $Re$ and and the 
asymptotic behavior of the most unstable mode in the large $Rm$ limit.  In particular it has been shown that: 
In a non-helical flow in an infinite domain the critical magnetic Reynolds number $Rm_c$ becomes a constant in the large $Re$ limit. A helical flow always results in dynamo in agreement with mean field predictions.
For thin layers for both helical and nonhelical flows the $Rm_c$ scales as a power-law of $Re$.
The growth-rate of fastest growing mode becomes independent of $Re$ and $Rm$ when their values are sufficiently large. The most unstable length scale in this limit scales linearly with the forcing length scale. Thus while the mean field predictions are valid, they are not expected to be dominant in the large $Rm$ limit.
\end{abstract}


\section{Introduction}
The dynamo instability caused by the motion of conducting fluids is used to explain the existence of magnetic fields in astrophysical objects. 
In many cases these objects are rotating rendering the resulting flow strongly anisotropic \citep{pedlosky2013geophysical, Izakov2013}. 
In rotating flows the Coriolis force suppresses the fluctuations along the axis of rotation as shown by the the Taylor-Proudmann theorem. Thus
very fast rotating flows become to some extent two-dimensionalized depending only on two spacial coordinates while retaining in some cases all three velocity components
depending on the boundary conditions. The two-dimensionalization of such flows have been shown in 
theoretical investigations \citep{Waleffe1993,Hopfinger1993,Scott2014}, 
numerical simulations      \citep{Hossain1994,Yeung1998,Smith1999,Chen2005,Thiele2009,Mininni2010,Yoshimatsu2011,Mininni2012,deusebio2014dimensional, alexakis2015rotating} and 
laboratory experiments     \citep{Sugihara2005, Staplehurst2008, Bokhoven2009, Yarom2013, Campagne2014direct, Gallet2014}. 
Recently a theoretical work showed that the flow becomes exactly two-dimensional provided that the rotation is above a critical value \citep{Gallet2015exact}. 
This allows one to consider the limit of infinite rotation which leads to a flow of the form $\left( u(x, y, t), v(x, y, t), w(x, y, t) \right)$. 
This flow is independent of the coordinate along the axis of rotation (from here on taken as the $z$-direction). 
These flows are referred in literature as $2.5D$ flows or $2+\epsilon$ model. 

Rotation is known to play an important role in dynamo instability \citep{1995lspdP,davidson2014dynamics}. A $2.5D$ flow is one of the simplest flow configuration that can result to dynamo
since a two-dimensional two-component flow does not give rise to dynamo instability \citep{Zeldovich:1957zl}.  
Thus, various studies have been performed in different limits. 
One of the first studies was by \cite{roberts1972dynamo} that considered  the dynamo instability of four different laminar time independent 2.5D flows. 
Time dependent but laminar 2.5D flows allow for the presence of chaos and thus pose a computationally tractable system to investigate the existence of the fast dynamos 
(dynamos whose growth rate remains finite in the high conductivity limit) that was investigated in \cite{galloway1992numerical},
as well as the behaviour of large scale dynamo action (alpha dynamos) in the same limit (see \cite{courvoisier2006alpha}). 
Studies of turbulent 2.5D flows were first studied to our knowledge by \cite{smith2004vortex}, where a helical forcing
was considered. In their configuration the inverse cascade of energy led to a large scale condensate which drove the dynamo instability. 
The role of these large scale coherent structures were further studied in \citep{tobias2008dynamo} where a differentiation between the scales 
responsible for the dynamo was made using spectral filters.

The present work focuses on turbulent 2.5D dynamos in the absence of large scales condensates.
Condensates form when an inverse cascade is present and there is no large scale dissipation mechanism 
to saturate the energy growth. Such a large scale dissipation mechanism is often provided by Ekman Friction \citep{pedlosky2013geophysical}
that leads to a linear damping. In its absence energy piles in the largest scales of the domain until it is balanced by
viscosity. This leads to the formation of very large amplitude vortices. However the turn-over time of the condensate vortices becomes comparable to the rotation period and the 
condition for quasi-two-dimensionalization is violated, with the flow becoming three dimensional again
\citep{Bartello1994,alexakis2015rotating}.  
For this reason we only consider here turbulent 2.5D flows in the presence of linear damping that 
limits the cascade to scales smaller than the domain size.

The study is based on numerical simulations of 2.5D turbulence in a two dimensional periodic box. 
Helical and non-helical flows are both considered. The focus is on covering a wide range of parameter space for both types of forcing. 
We describe the system in detail in Section \ref{Section:Two} and discuss the hydrodynamic cascades that happen in this set-up in Section \ref{Section:Three}. 
The results for the helical forcing are presented in Section \ref{Section:Four} and for the non-helical forcing in Section \ref{Section:Five}. The critical 
magnetic Reynolds number is discussed in section \ref{Section:Six}. The dependence of the dynamo instability with respect to the forcing length-scale is discussed in
Section \ref{Section:Seven}. We present our conclusions  in Section \ref{Section:Eight}.

\section{Governing equation} \label{Section:Two}

We consider a $2.5D$ flow in a box of size $[2 \pi L, 2 \pi L, H]$ with the height $H$ being along the invariant direction $z$. 
The equations governing the velocity field ${\bf u} = {\bf u}_{_{2D}} + u_z \hat{\bf e}_z = \nabla \times \left( \psi \hat{\bf e}_z \right) + u_z \hat{\bf e}_z$ are, 
\begin{eqnarray}
\partial_t \Delta \psi + \left( {\nabla \times \psi \hat{\bf e}_z} \right) \cdot {\nabla} \, \Delta \psi = &  \, \nu \,  \Delta^2 \psi - \nu_{_h} \, \Delta \psi + \Delta f_{\psi} \nonumber \\
\partial_t {u_z} + \left( {\nabla \times \psi \hat{\bf e}_z} \right) \cdot {\nabla} u_z \, = &  \, \nu \,  \Delta  {u}_z + f_z
\end{eqnarray}
where $\nu$ is the small scale dissipation coefficient, $\nu_{_{h}}$ is the large scale dissipation coefficient. The vertical velocity $u_z$ is advected as a passive scalar. We force the velocity fields with the forcing: $f_{\psi}, f_z$. Two forcing functions are used to study this problem, one with mean helicity and the other without any mean helicity. 
More precisely the forcing functions are, $f_{\psi} = f_z = \cos \left( k_f x \right) + \sin \left( k_f y \right)$ for the helical case and $f_{\psi} = \cos \left( k_f x \right) + \sin \left( k_f y \right), f_z = \sin \left( k_f y \right) + \cos \left( k_f x \right) $ for the nonhelical case. It is easy to note that for the helical case the helicity of the forcing given by $\left\langle f_z \Delta f_{\psi} \right\rangle \neq 0$ whereas for the nonhelical case it is zero.  

Due to the invariance in the $z$ direction the magnetic field can be decomposed into Fourier modes in $z$, ${\bf B} = {\bf b}(x, y, t) \, exp(i k_z z)$. Each mode evolves independently and is governed by the induction equation, 
\begin{eqnarray}
\partial_t {\bf b} + \left( \nabla \times \psi \hat{\bf e}_z \right) \cdot \nabla \, {\bf b} + u_z i k_z {\bf b} = {\bf b} \cdot \nabla \, \left( \nabla \times \psi \hat{e}_z \right) + \eta \, \left( \Delta - k_z^2 \right) {\bf b}
\end{eqnarray} 
where $\eta$ is the magnetic diffusion. The divergence free condition $\nabla \cdot {\bf B} = 0$ for each magnetic mode gives, 
\begin{eqnarray}
\partial_x b_x(x, y, t) + \partial_y b_y(x, y, t) = - i k_z b_z(x, y, t).
\end{eqnarray}
The non-dimensional control parameters of this system are the $Re = \left\langle |{\bf u}|^2 \right\rangle^{1/2} L/\nu$ the fluid Reynolds number, $Rm = \left\langle |{\bf u}|^2 \right\rangle^{1/2} L/\eta$ the magnetic Reynolds number, $k_f L$ the forcing wavenumber. Given the set of non-dimensional parameters we look for the range of modes $k_z L$ that become unstable. 

The equations are solved numerically on a double periodic domain of size $[2 \pi L, 2 \pi L]$ using a standard pseudo-spectral scheme and a Runge-Kutta fourth order scheme for time integration (see \cite{Gomez05}). The initial condition for both the magnetic and the kinetic field is sum of a few Fourier modes with random phases. Initially a hydrodynamic steady state is obtained by solving only the hydrodynamic equations at a particular $Re, k_f L$. With this steady state the dynamo simulation is begun with a seed magnetic field and evolving both the velocity and the magnetic field. The magnetic field starts to grow or decay depending on the control parameters in the system. We define the growth rate of the magnetic field as,
\begin{eqnarray}
\gamma = \lim_{t\rightarrow \infty} \frac{1}{2t} log \frac{\left\langle |{\bf b}|^2(t) \right\rangle}{\left\langle |{\bf b}|^2(0) \right\rangle}
\end{eqnarray} 
as a function of the non-dimensional parameters $Re, Rm, k_z L, k_f L$. A table of runs is shown in table \ref{Table:Allruns} indicating the range of values of each parameters examined. 
\begin{table}
  \caption{Range of values of each parameter explored in the DNS. $N$ is the numerical resolution in each direction and $T$ is the typical eddy turn over time over which the growth rate is calculated.}
   \centering
   \begin{tabular}{c|ccc} 
		Case   \hspace{10mm}         &    \hspace{10mm} A1  \hspace{10mm} &       \hspace{10mm} A2 \hspace{10mm}  & \hspace{10mm}    A3  \hspace{10mm}     \\ \hline
    		 $k_f L$   \hspace{10mm}   &  4               &    8              &   16             \\ 
         $N$   \hspace{10mm}     & $[256, 2048]$    &   $[512, 2048]$   &   $[512, 2048]$  \\  
         $Re$   \hspace{10mm}    & $[0.5, 1200]$     &   $81,92$         &   $91,97$   \\
         $Rm$   \hspace{10mm}    & $[0.1,2000]$    &  $[0.5,1000]$     &  $[0.5,1000]$ \\
         $T$    \hspace{10mm}    & $[300,2000]$     &  $[300,600]$      &   $[300,600]$   \\
    \end{tabular}
  \label{Table:Allruns}
\end{table}

\section{Hydrodynamic cascades} \label{Section:Three}

We first describe the hydrodynamic structure of the flow. The quantities conserved by the nonlinearities in the hydrodynamic equations are, the enstrophy in $x-y$ plane $\Omega = \left\langle \omega_z^2 \right\rangle$ with $\omega_z = - \Delta \psi$ and the angular brackets $\left\langle \cdot \right\rangle$ denote spatial average, the energy in $x-y$ plane, $E_{_{2D}} = \left\langle {\bf u}_{_{2D}} \cdot {\bf u}_{_{2D}} \right\rangle/2$, the energy of the $z$ component of velocity $E_{_Z} = \left\langle {u}_z^2 \right\rangle/2$ and the helicity $H = \left\langle u_z \, \omega_z \right\rangle$. For a more detailed discussion on the invariants see \cite{smith2004vortex}. For sufficiently small viscosity $\nu$ and damping $\nu_h$ the conserved quantities cascade
either to the small or the large scales. 
For the $2D$ flow we have a forward cascade of enstrophy $\Omega$ and an inverse cascade of energy $E_{_{2D}}$. 
The $E_z$ has a forward cascade since $u_z$ is advected like a passive scalar. 
We mention that even when the forcing is with maximum helicity $f_z \propto f_{\psi}$ the governing equation for $u_z$ is not the same as the vorticity $\omega_z$ due to the presence of large scalar damping. 
Helicity cascades to small scales since both $E_z, \Omega$ cascade to small scales. Between the forcing scale and the dissipation scale there exists a range of scales (the inertial range) where the energy spectra have a power law behaviour. The exponents of these power laws are determined by the cascading quantities in the classical Kolmogorov approach. The exponents in this setup, for $E_{_{2D}}$, is $-3$ in the scales smaller than the forcing scale due to enstrophy cascade and $-5/3$ in the scales larger than the forcing scale due to the inverse energy cascade. Similarly for the spectra of $E_z$ we have $-1$ in the scales smaller than the forcing scale due to the forward cascade of the passive scalar $E_z$ like the spectra of the variance of a passive scalar \citep{batchelor1959small}. Since there are no inverse cascade for $u_z$ we expect an equipartition exponent of $+1$ at scales larger than the forcing scale. 

\begin{figure}
\begin{center}
\includegraphics[scale=0.13]{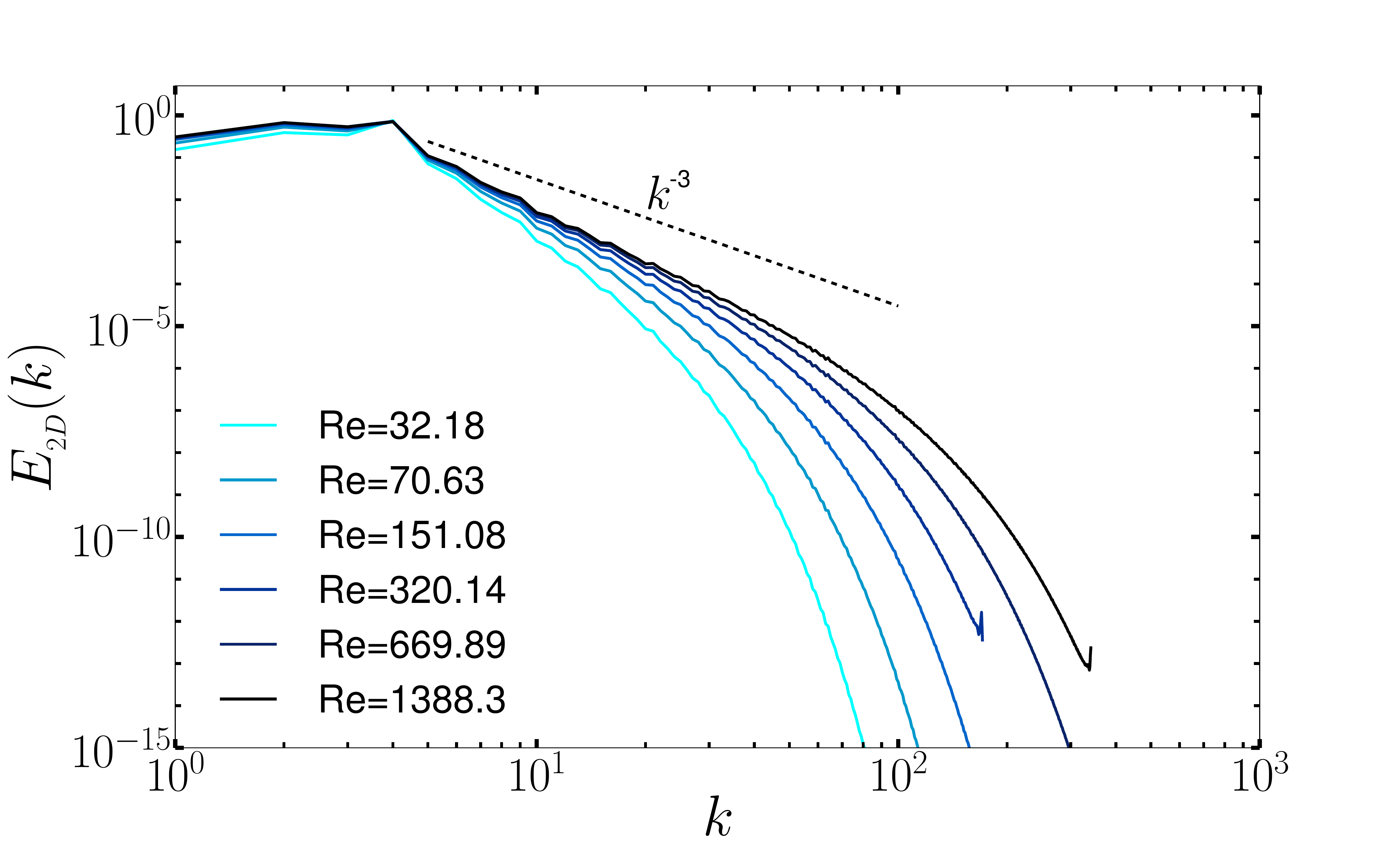}
\includegraphics[scale=0.13]{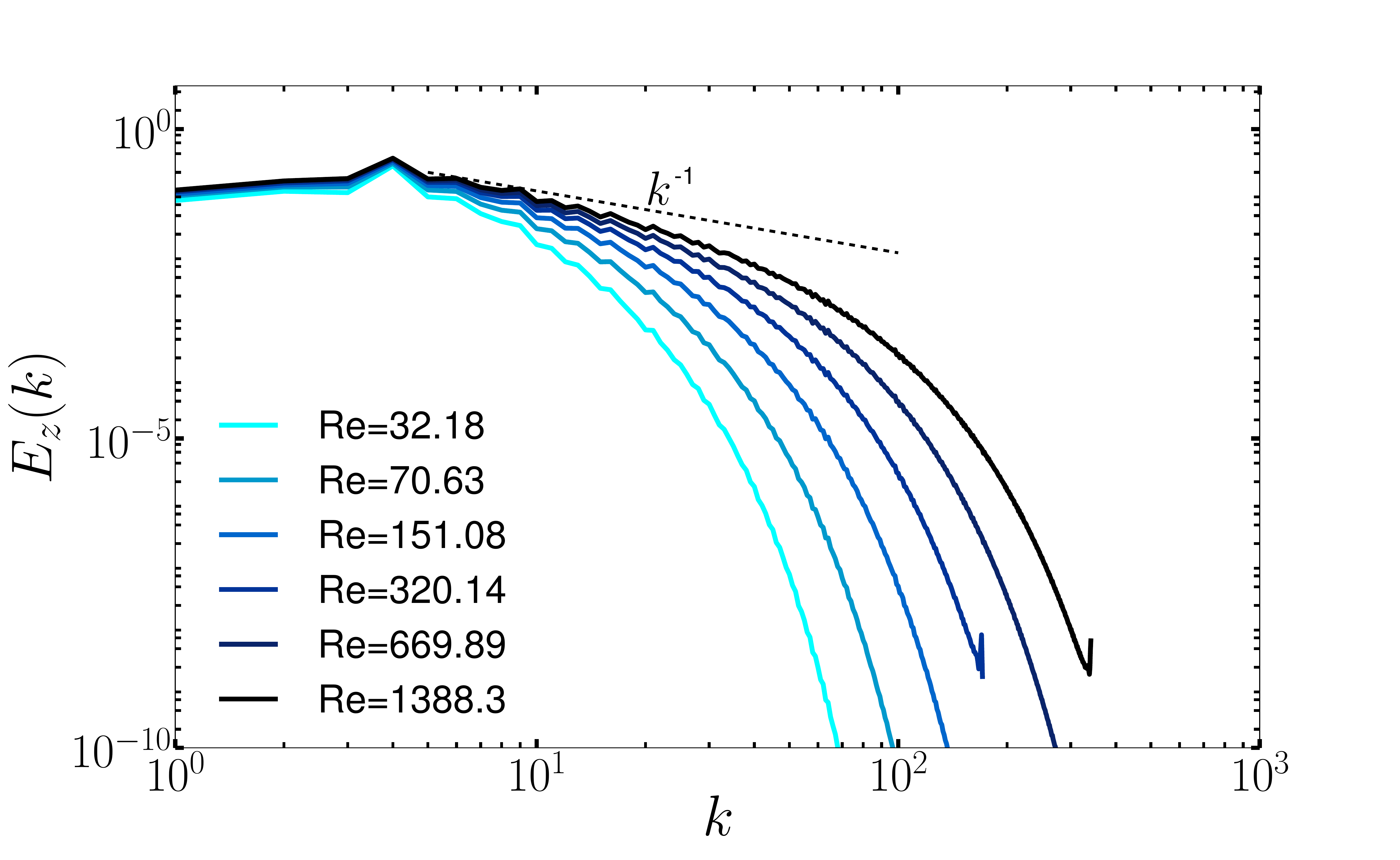}
\end{center}
\caption{Plot shows the spectra of the $2D$ kinetic energy $E_{_{2D}} (k)$ and the spectra of the vertical velocity $E_{z} (k)$ for different values of $Re$ mentioned in the legend. The spectra correspond to nonhelical forcing case.}
\label{Fig:Spectra_Ek}
\end{figure}

\begin{figure}
\begin{center}
\includegraphics[scale=0.13]{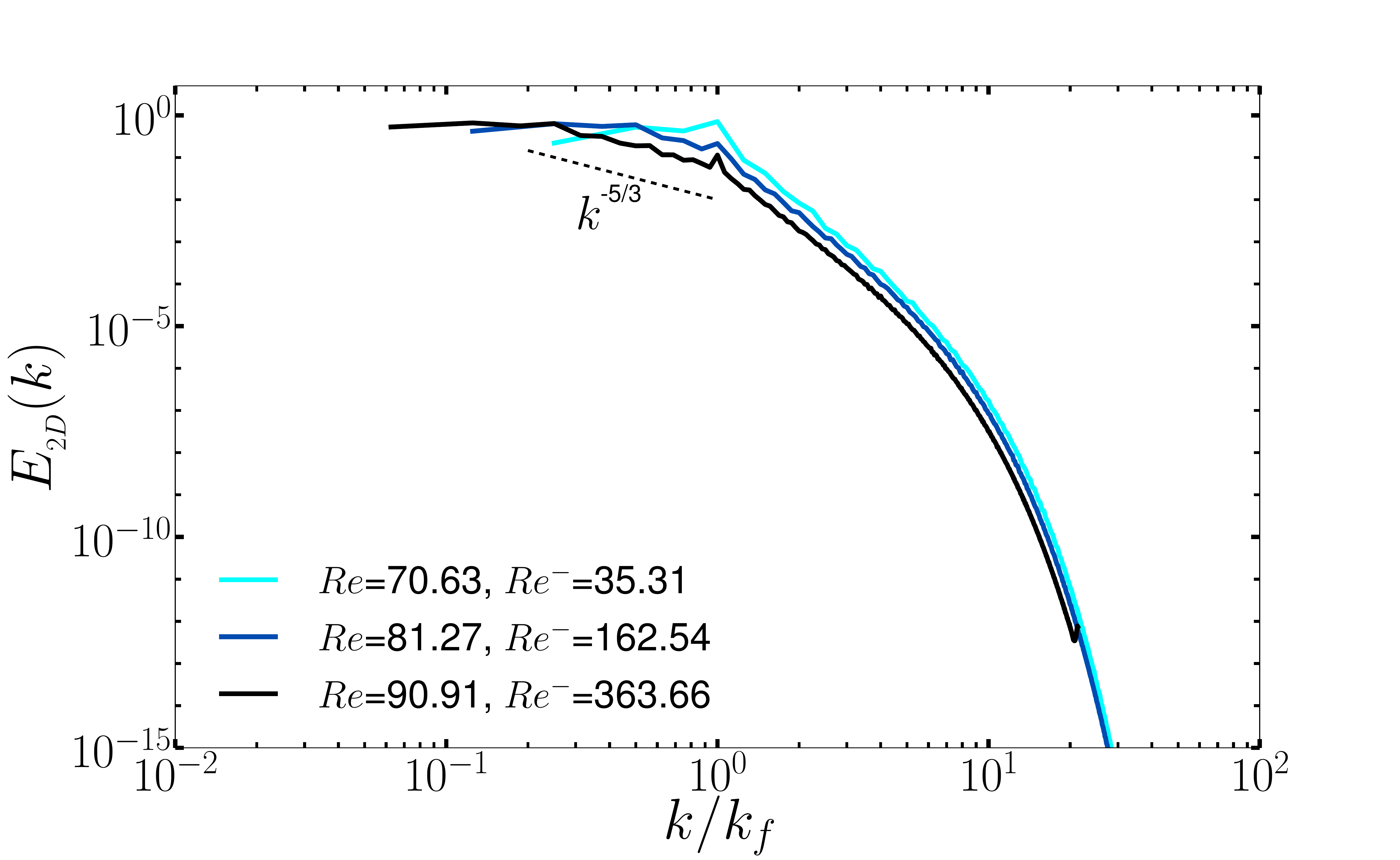}
\includegraphics[scale=0.13]{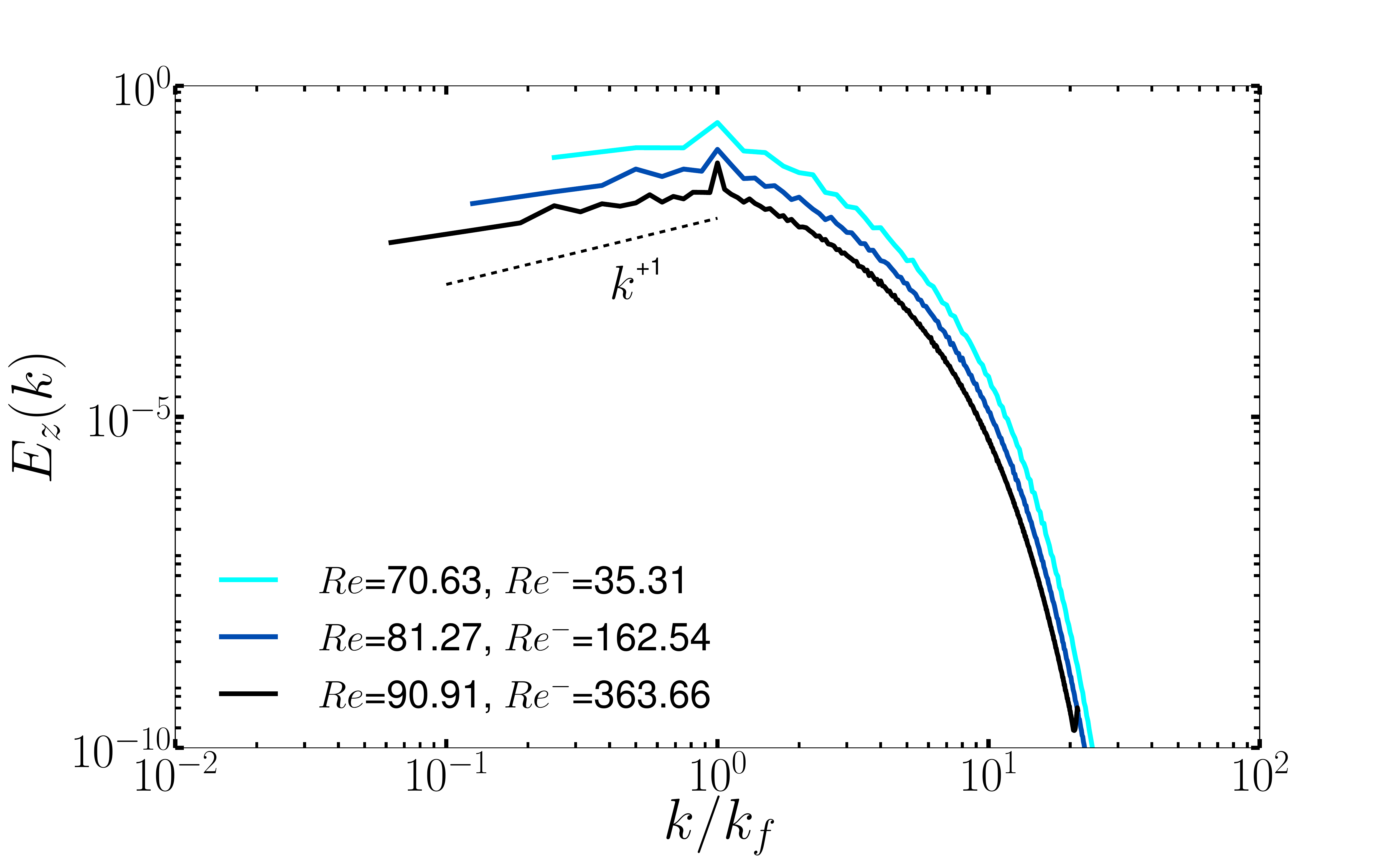}
\end{center}
\caption{Plot shows the spectra of the $2D$ kinetic energy $E_{_{2D}} (k)$ and the spectra of the vertical velocity $E_{z}$ as a function of the rescaled wavenumber $k/k_f$ for different values of $Re$ and $k_f \, L$ mentioned in the legend. The spectra correspond to the nonhelical forcing case.}
\label{Fig:diff_kf}
\end{figure}

Figure \ref{Fig:Spectra_Ek} shows the spectra $E_{_{2D}}$ and $E_z$ for different values of $Re$ for nonhelical forcing. The spectra of the helical forcing case are very similar to the spectra of the flows with nonhelical forcing so they are not shown here. The figure shows that the exponents of $E_{_{2D}}$ and $E_z$ in the forward cascade change as we increase the $Re$. As shown in \cite{boffetta2007energy} the exponent for the energy spectra in the small scales tend to the expected value of $-3$ as the $Re$ becomes large. In their study they went upto $32768^2$ to get the expected $k^{-3}$ spectrum. In this work since the focus is on the dynamo effect the simulations are done only upto $2048^2$, thus the exponent in the spectra is less than $-3$. Figure \ref{Fig:diff_kf} shows the spectra $E_{_{2D}}$ and $E_z$ as $k_f \, L$ is varied for the nonhelical forcing. Due to the presence of an inverse cascade the energy spectra form a $k^{-5/3}$ for scales larger than the forcing scale. While for the vertical velocity spectra the large scales form an equipartition spectrum of $k^{+1}$. The inverse cascade of energy is dissipated by the friction at large scales which inhibits the formation of a large scale condensate. 

The transfer of energy to the magnetic field from the kinetic field in the dynamo problem is related to the shear of the velocity field. In $2D$ turbulence, the shear $S_{_{2D}}^{\ell}$ in ${\bf u}_{_{2D}}$ at a scale $\ell$ can be estimated by $S_{_{2D}}^{\ell} \propto u_{_{2D}}^{\ell}/\ell$ where $u_{_{2D}}^{\ell}$ is the amplitude of the ${\bf u}_{_{2D}}$ at a scale $\ell$. For $2D$ turbulence $S_{_{2D}}^{\ell}$ is same at all scales between the forcing and the small scale dissipation since $u_{_{2D}}^{\ell} \sim \ell$. Thus for any $\ell_f>\ell>\ell_{\nu}$ we have $S_{_{2D}}^{\ell} \ell_f/u_f \sim \mathcal{O}(1)$. This is strictly true for a spectra of $k^{-3}$, which is seen at very large $Re$. Since most of the study presented here is with an exponent less than $-3$ the shear $S_{_{2D}}$ is dominated at the forcing scale $k_f$. For the vertical velocity field the shear can be estimated by $S_z^{\ell} \propto u_z^{\ell}/\ell$, where $u_z^{\ell}$ is the magnitude of $u_z$ at scale $\ell$. It is dominated at the smallest scales and we have $S_z^{\ell_{\nu}} \ell_f/u_f \sim Re^{1/2}$. Thus shear is dominated at forcing scale for ${\bf u}_{_{2D}}$ while for $u_z$ it is dominated at the viscous scales. However the dynamo instability requires the presence of both $S_z$ and $S_{\ell}$. Thus we can not a priori determine which scales will be responsible for dynamo action. We remark that the dominant shear scales present here differ from the condensate regime where all the shear is dominated at the scale of the box. 
\section{Helical forcing} \label{Section:Four}

\subsection{Dependence of $\gamma$ on $k_z$} 

\begin{figure}
\begin{center}
\includegraphics[scale=0.2]{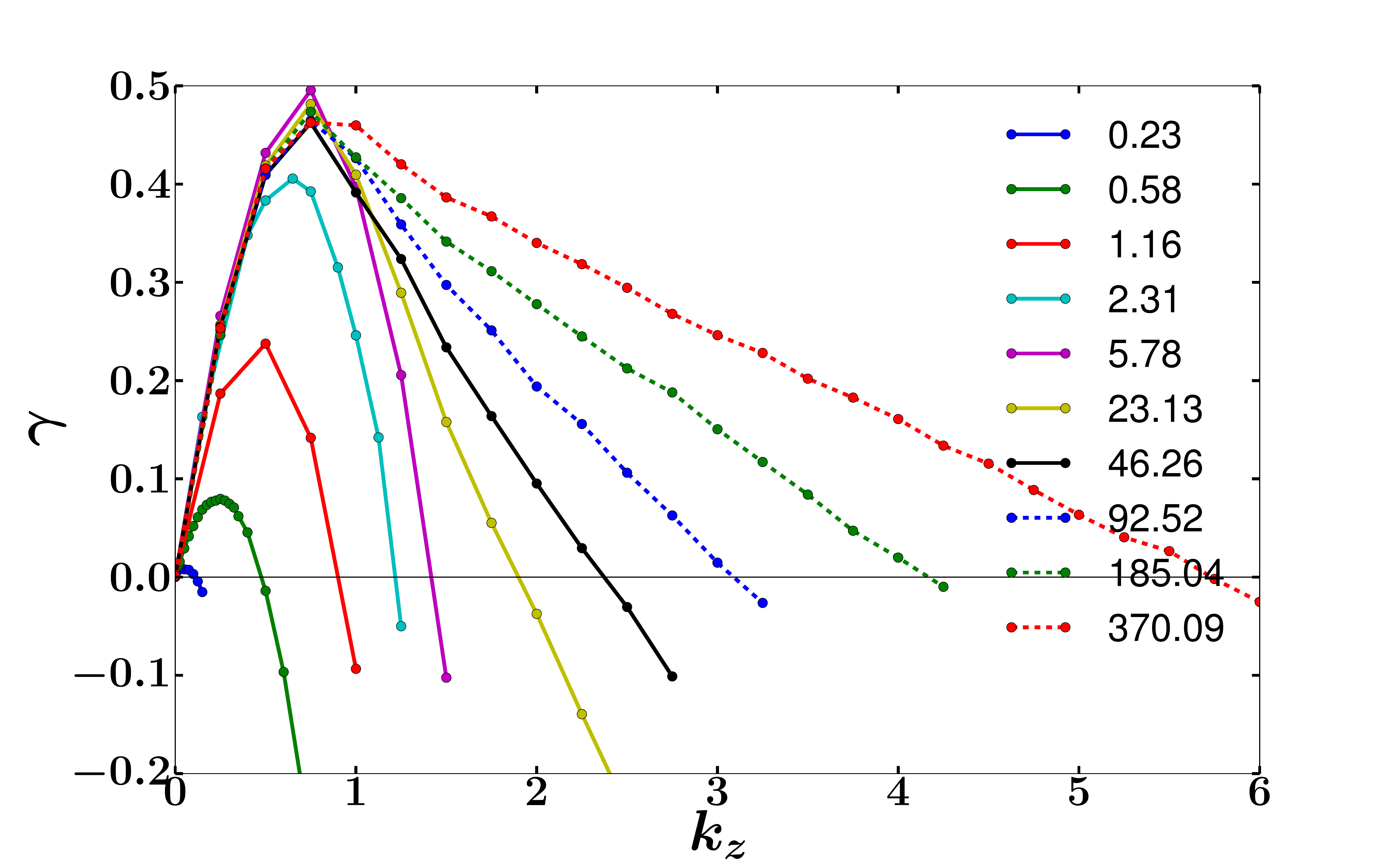}
\end{center}
\caption{Plot shows the growth rate $\gamma$ as a function of $k_z$ for the helical forcing case for different values of $Rm$ mentioned in the legend for a $Re \approx 46$.}
\label{Fig:gammavskzhel}
\end{figure}

We first focus on the helical forcing, the laminar case of which corresponds to the case studied by \cite{roberts1972dynamo}. Figure \ref{Fig:gammavskzhel} shows the growth rate $\gamma$ as a function of $k_z$ for different values of $Rm$ that are mentioned in the legend and for a fixed $Re \approx 46$. 
The number of unstable $k_z$ modes increases as we increase $Rm$  as has been observed in other laminar and turbulent studies \cite{roberts1972dynamo,tobias2008dynamo, smith2004vortex}. As we increase $Rm$ the growth rates for small $k_z \sim \mathcal{O}(1)$ saturates. 

\begin{figure}
\begin{center}
\includegraphics[scale=0.13]{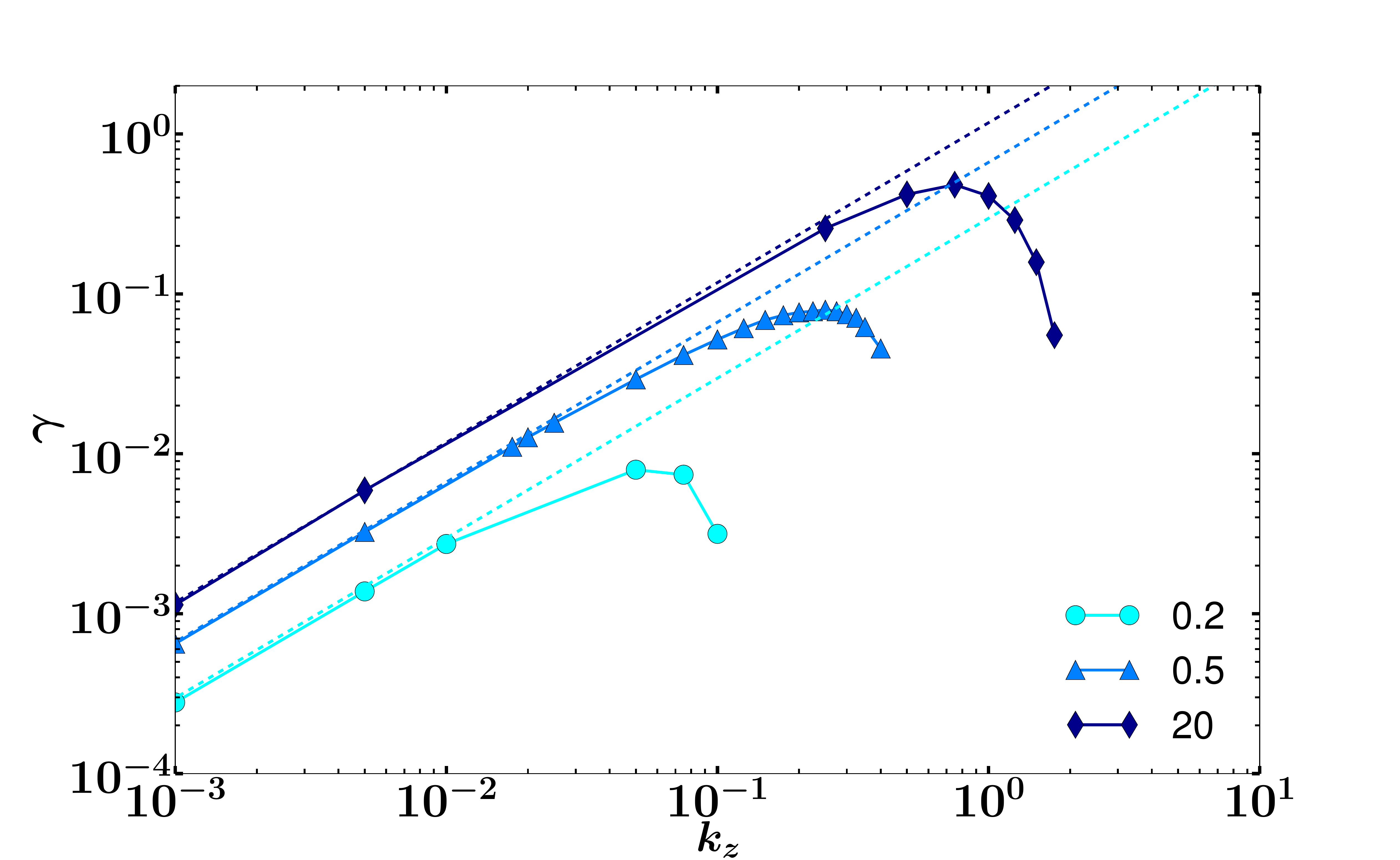}
\includegraphics[scale=0.13]{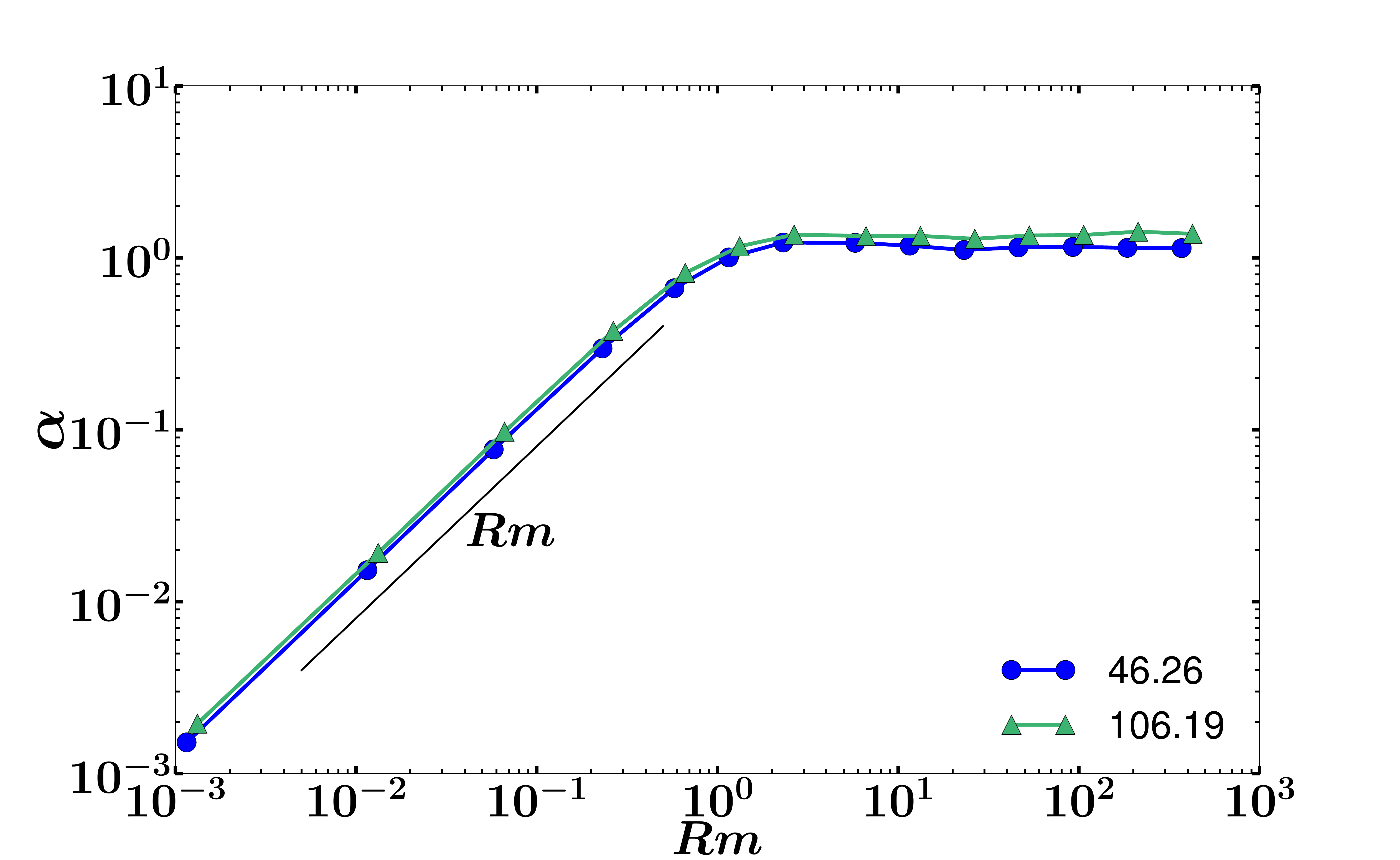}
\end{center}
\caption{The figure on the left shows the growth rate $\gamma$ as a function of $k_z$ in log-log scale. The corresponding $\alpha$ values are shown by the dotted straight lines at values of $Rm$ mentioned in the legend. The figure on the right shows $\alpha$ as a function of $Rm$ for two different $Re$ mentioned in the legend.}
\label{Fig:alphacoeff_growth}
\end{figure}

There are dynamo unstable modes for all values of $Rm$, but the range of unstable modes become smaller as $Rm$ is reduced. This can be attributed to the $\alpha$-effect which is a mean field effect that can amplify the magnetic field at arbitrarily large scales. 
In the mean field description the large scale magnetic field $\bf \overline{B}$ obeys the equation
\begin{align}
\partial_t \bf {\overline{B}} = \nabla \times (\alpha  \overline{B}) + \eta_{_T} \Delta \overline{B}
\end{align} \label{Eqn:alphaeq}
where $\alpha$ is in general a tensor and $\eta_{_T}$ is the turbulent diffusivity.
For isotropic flows the diagonal terms in the $\alpha$ tensor are equal and are responsible for the dynamo effect. 
They can be calculated numerically by imposing a uniform magnetic field ${\bf B}_0$ and measuring the induced field ${\bf b}$, (see \cite{courvoisier2006alpha}). 
\begin{align}
{\bf \alpha} \cdot {\bf B}_0 & = \left\langle {\bf u} \times {\bf b} \right\rangle \label{Eqn:alphacal} \\
\partial_t {\bf b} + {\bf u} \cdot \nabla {\bf b} & = {\bf b} \cdot \nabla {\bf u} + {\bf B}_0 \cdot \nabla {\bf u} + \eta \Delta {\bf b} \label{Eqn:alphaeqn}
\end{align}
In the small $Rm$ limit, $\eta_{_T}=\eta$ and the $\alpha$ coefficient can be calculated analytically (see \cite{childress1969class,gilbert2003dynamo}) leading to the scaling $\alpha \sim u \, Rm$.
In either case the resulting growth rate for the problem at hand is given by
\begin{align}
\gamma = \alpha \, k_z - \eta_{_T} \, k_z^2. 
\end{align} \label{Eqn:gammaalpha}
The left panel of figure \ref{Fig:alphacoeff_growth}, shows the $\gamma-k_z$ curve in log-log scale with the straight lines indicating the linear scaling $\alpha \, k_z$ with $\alpha$ calculated from
equations \ref{Eqn:alphacal}, \ref{Eqn:alphaeqn}. This demonstrates that the behaviour of $\gamma$ in the small $k_z$ limit is described well by the $\alpha$-effect. The right panel of figure \ref{Fig:alphacoeff_growth} shows the
dependence of $\alpha$ as a function of $Rm$ for two different $Re$. For a turbulent flow and for small $Rm$ we expect the $\alpha$ coefficient to scale like $\alpha \sim u \, Rm$, see 
\cite{gilbert2003dynamo}, which is captured well by the numerical data. For large $Rm$ the $\alpha$ value saturates to a constant of the same order as the velocity field. 
This is different from what has been observed in chaotic flows in \cite{courvoisier2006alpha}, where the $\alpha$ coefficient varies rapidly as one increases $Rm$. 

\begin{figure}
\begin{center}
\includegraphics[scale=0.15]{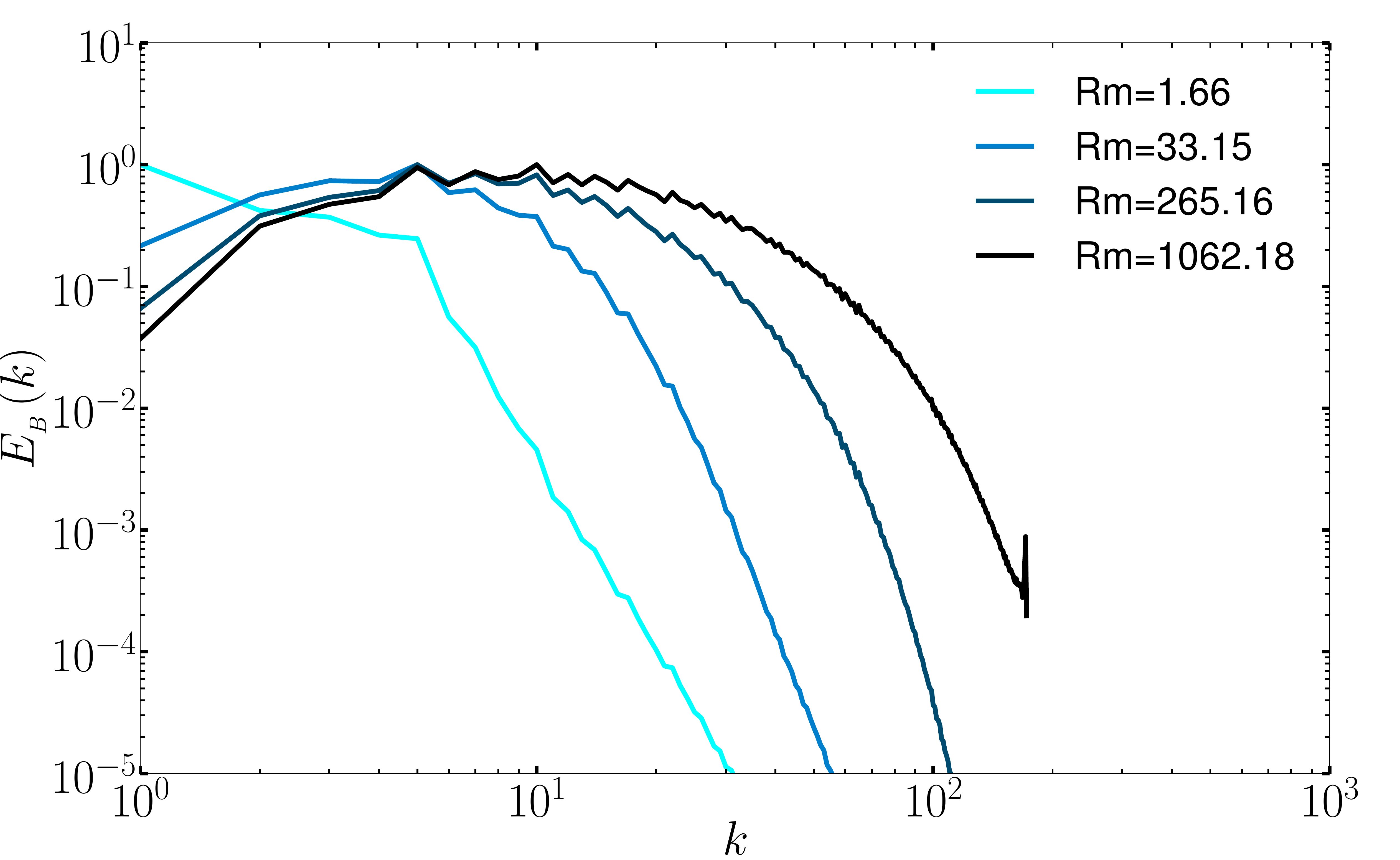}
\end{center}
\caption{Plot shows the magnetic energy spectra $E_{_B}(k)$ as a function of the wavenumber $k$ for different $Rm$ shown in the legend. These correspond to a Reynolds number $Re \approx 530$ and to the helical forcing case.}
\label{Fig:Magspec}
\end{figure}

Figure \ref{Fig:Magspec} shows the total magnetic energy spectra $E_{_B}(k)$ for different values of $Rm$ and a fixed $k_z = 0.25$ and $Re \approx 530$. When the $\alpha$ effect is more pronounced, the magnetic spectra is concentrated at large scales. This occurs in the small $Rm$ limit. For large $Rm$ the magnetic energy spectra becomes more concentrated towards smaller scales. 

\subsection{$\gamma_{max}$ and $k_z^c$} 
\begin{figure}
\begin{center}
\includegraphics[scale=0.13]{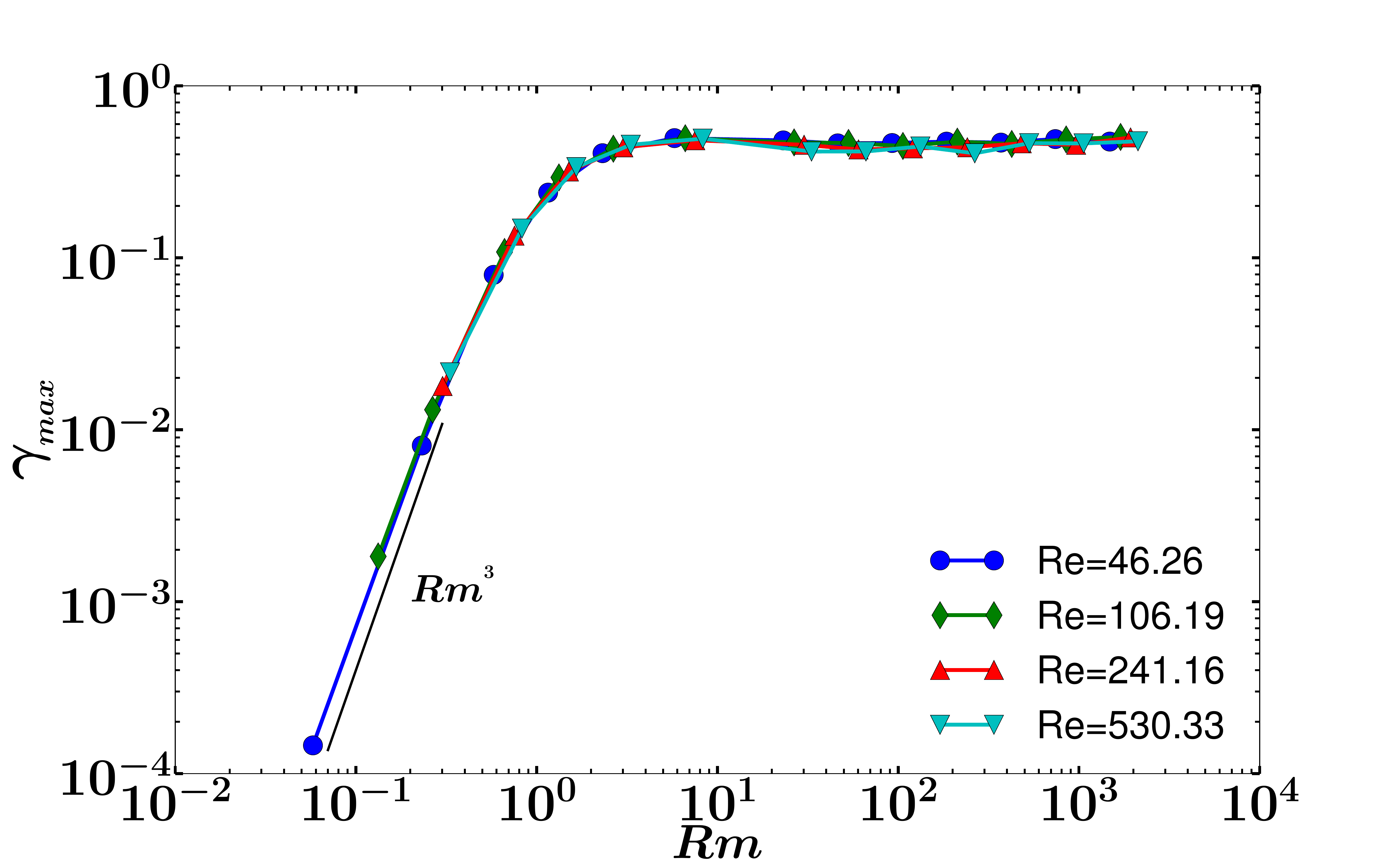}
\includegraphics[scale=0.13]{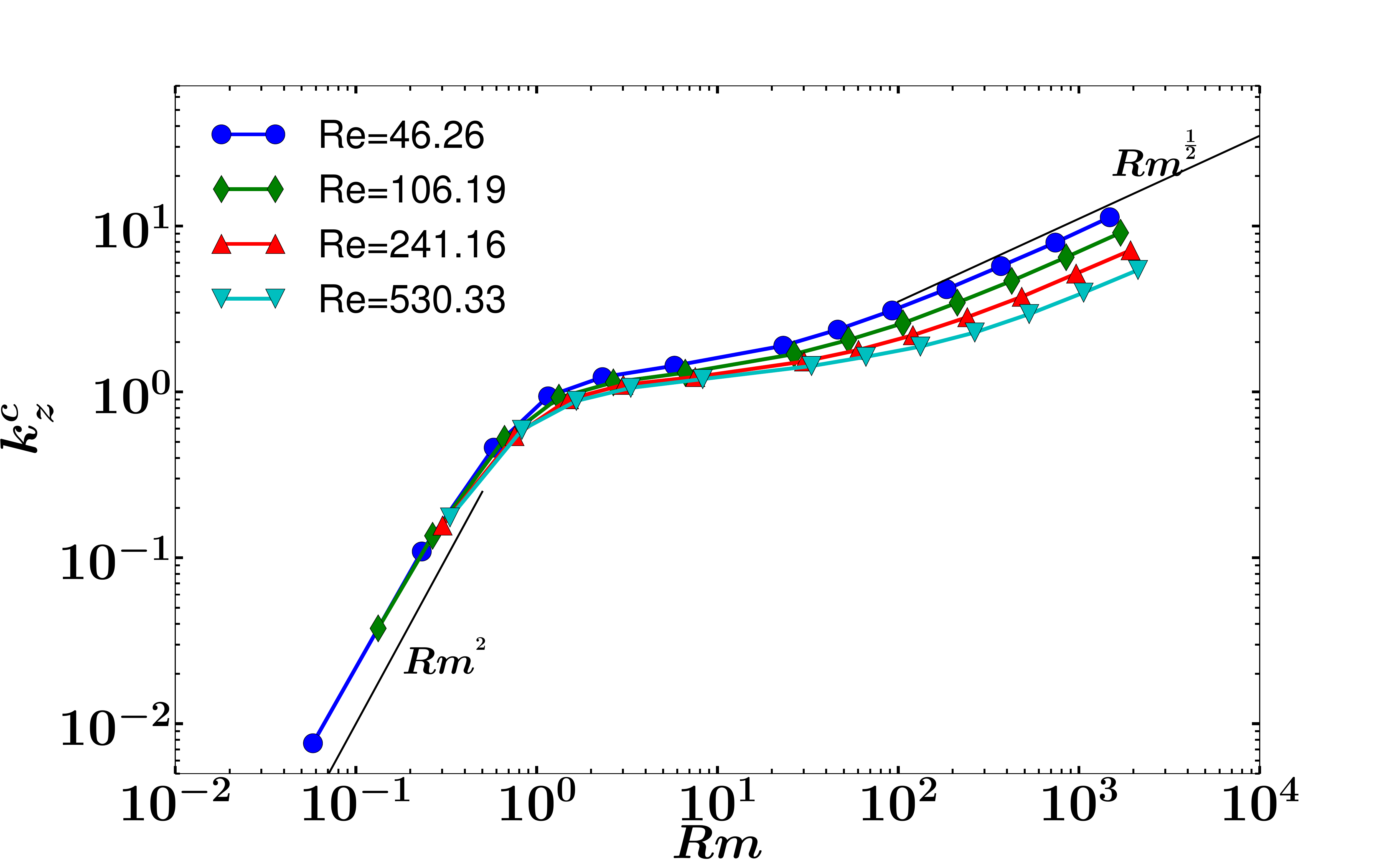}
\end{center}
\caption{Figure shows $\gamma_{max}$ on the left and $k_z^c$ on the right as a function of $Rm$ for different values of $Re$ mentioned in the legend for the helical forcing case.}
\label{Fig:gamma_kzchel}
\end{figure}

To quantify the behaviour of $\gamma$ as we change both $Re$ and $Rm$ we define two quantities $\gamma_{max}$ and $k_z^c$ which characterize the curves shown in figure \ref{Fig:gammavskzhel}. $\gamma_{max}$ is the maximum growth rate for a given $Re, Rm$ whereas $k_z^c$ is the largest $k_z$ that is dynamo unstable for a given $Re, Rm$. Figure \ref{Fig:gamma_kzchel} shows $\gamma_{max}$ and $k_z^c$ as functions of $Rm$ for different values of $Re$. It can be seen that $\gamma_{max}$ is independent of $Re$ and becomes saturated for $Rm \sim \mathcal{O}(1)$. In the small $Rm$ limit the behaviour of $\gamma_{max}$ is governed by the $\alpha$-effect, which gives a scaling $\gamma_{max} \propto Rm^3$ obtained by finding the maximum of equation \ref{Eqn:gammaalpha}. For large $Rm$ the $\gamma_{max}$ remains a constant thus it is a fast dynamo. The most unstable length scale is in between the forcing scale and scale of the box and not in the small $k_z$ region.

In the plot of $k_z^c$ in the small $Rm$ limit the behaviour is dominated by the $\alpha$-effect where $k_z^c \propto Rm^2$ obtained from equation \ref{Eqn:gammaalpha}. In this limit the behaviour of $k_z^c$ does not depend on $Re$ since $k_z^c = c Rm^{2}$ with $c$ being independent of $Re$. For large values of $Rm$ we see the scaling $k_z^c \propto Rm^{1/2}$ which can be obtained by balancing the ohmic dissipation with the stretching term. We can also see a clear decrease with the increase of $Re$ which will be discussed in section \ref{Sec:Finitedomain}. 

\section{Nonhelical forcing} \label{Section:Five}

\subsection{Dependence of $\gamma$ on $k_z$}

\begin{figure}
\begin{center}
\includegraphics[scale=0.2]{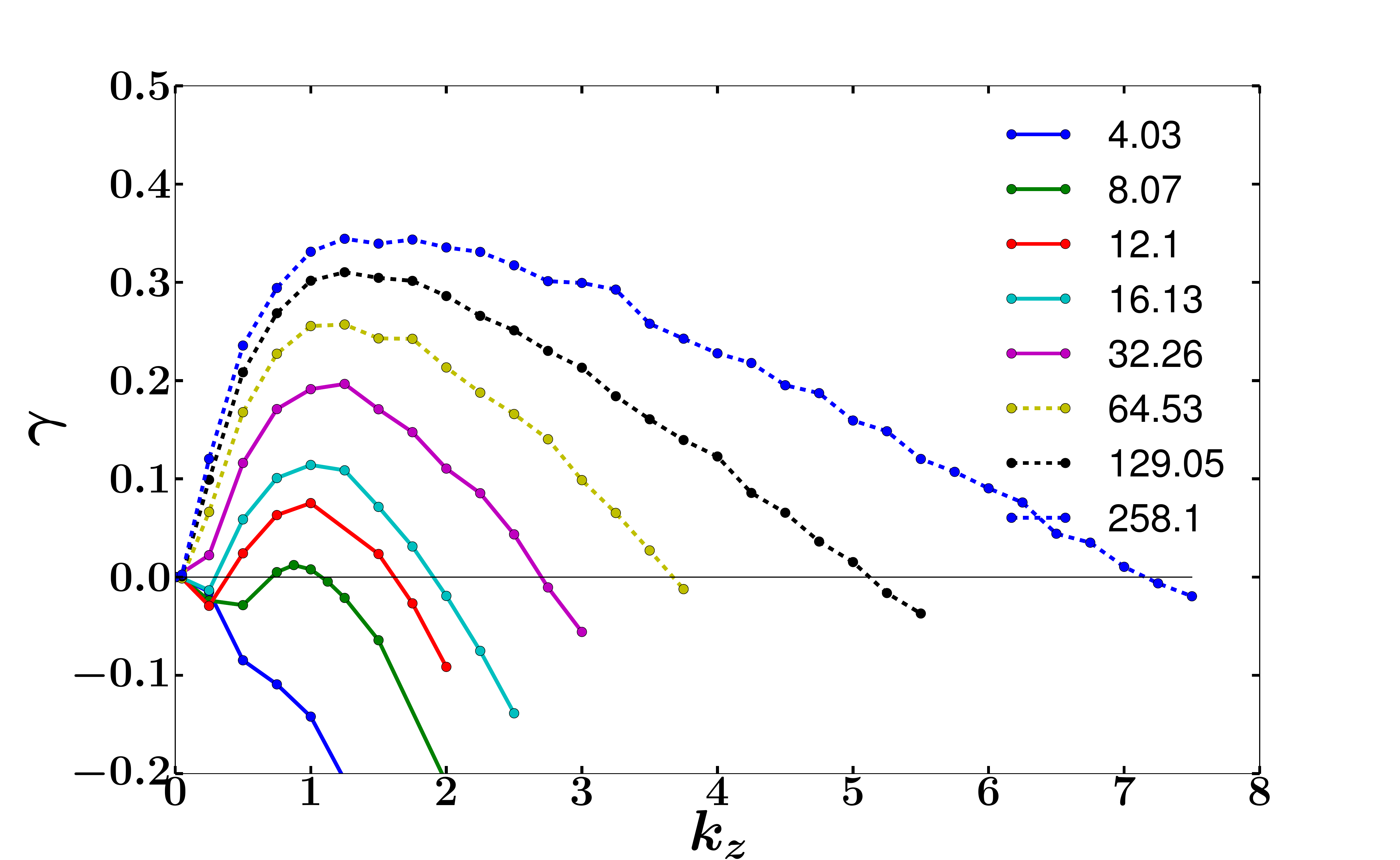}
\end{center}
\caption{Plot shows the growth rate $\gamma$ as a function of $k_z$ for different values of $Rm$ mentioned in the legend for a $Re \approx 32$. The curves correspond to the nonhelical forcing case. }
\label{Fig:gammavskznonhel}
\end{figure}

The growth rate $\gamma$ is shown as a function of $k_z$ for different values of $Rm$ in figure \ref{Fig:gammavskznonhel}. Unlike the helical case, there is no dynamo for small $Rm$ due to the absence of a mean-field $\alpha$-effect. For sufficiently large $Rm$ dynamo instability occurs with the magnetic spectra concentrated in the small scales similar to the large $Rm$ case of the helical forcing shown in figure \ref{Fig:Magspec}. As $Rm$ is increased the number of unstable modes increase as the ohmic dissipation becomes smaller. 
%
\subsection{$\gamma_{max}$ and $k_z^c$}
\begin{figure}
\begin{center}
\includegraphics[scale=0.13]{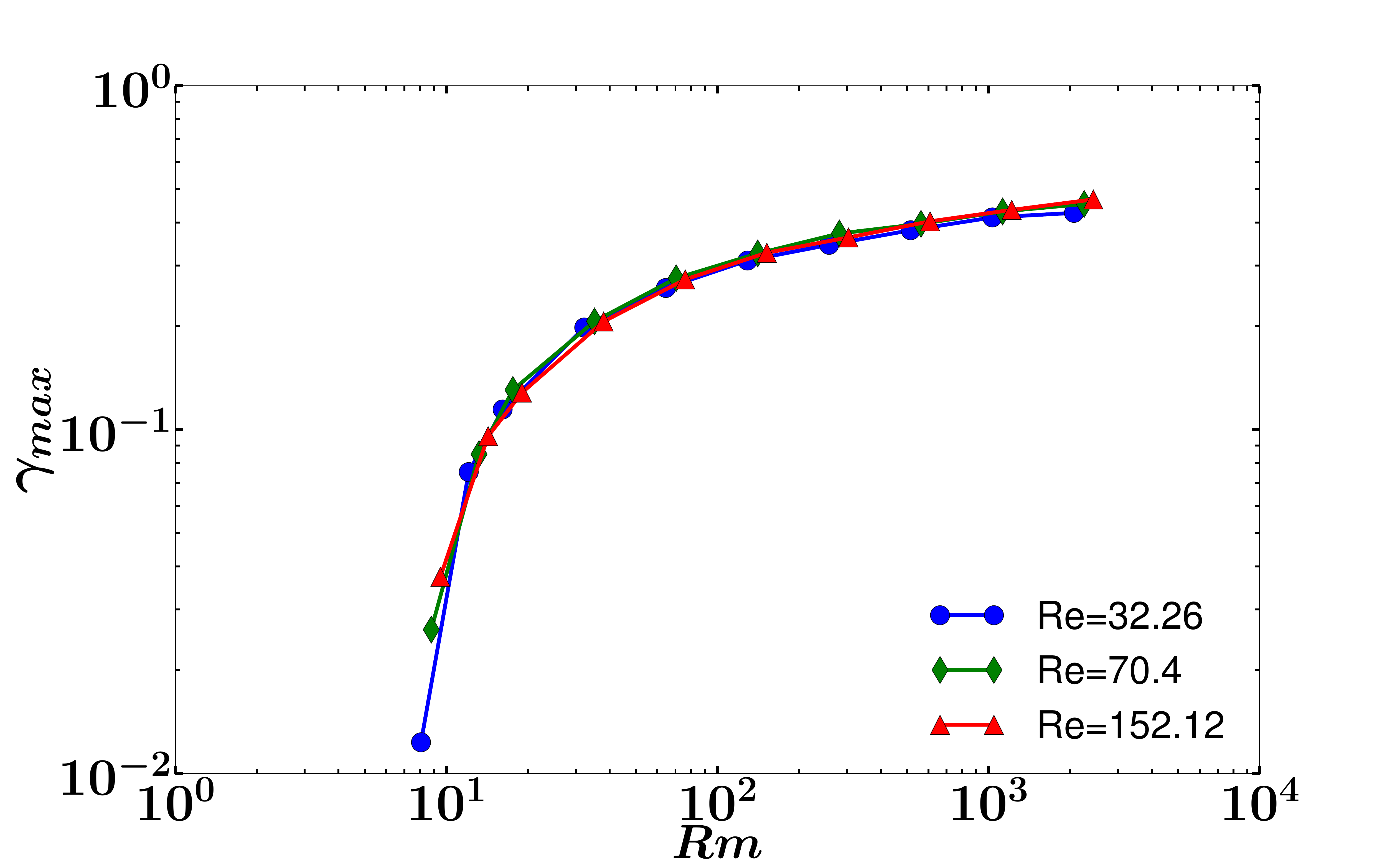}
\includegraphics[scale=0.13]{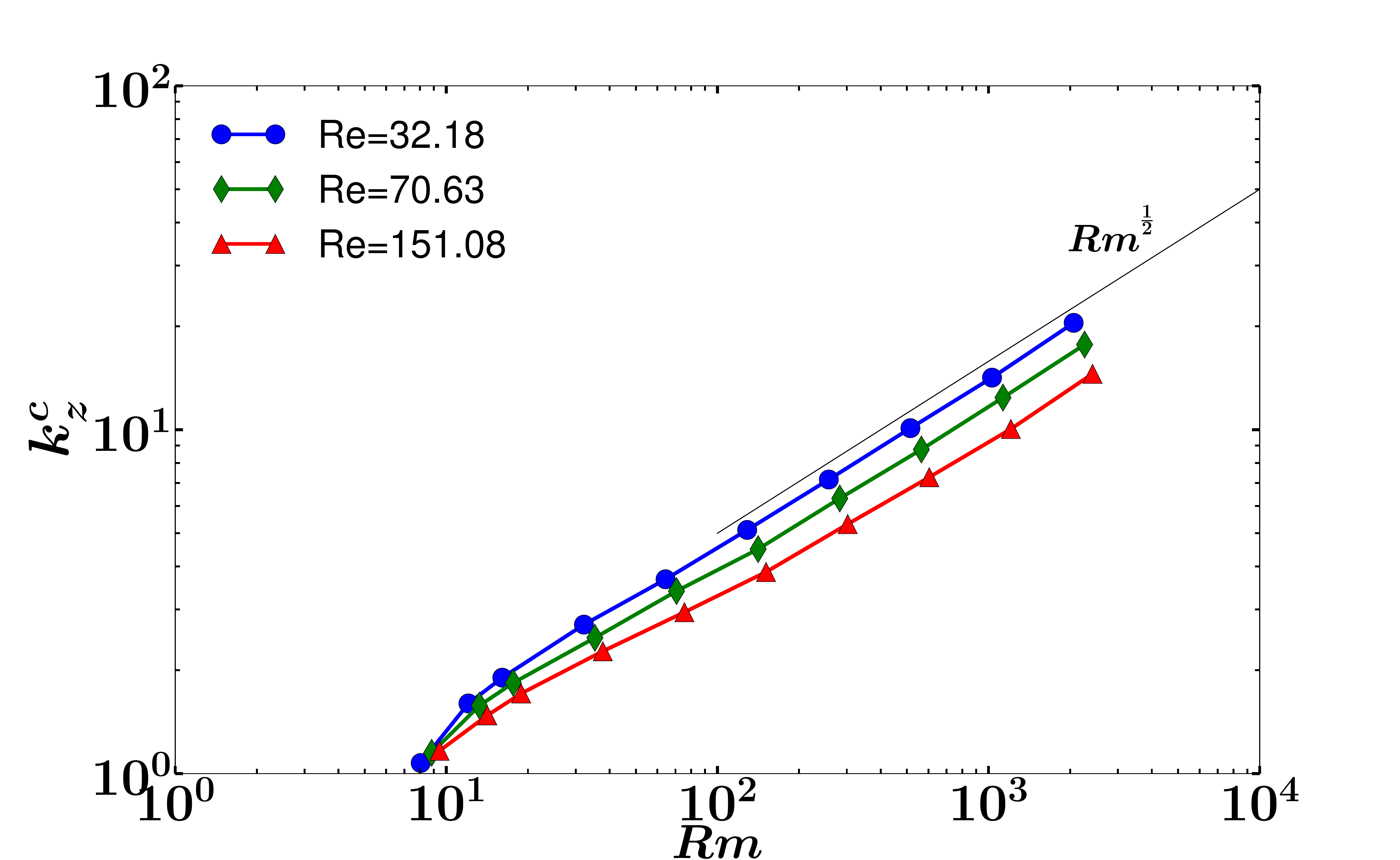}
\end{center}
\caption{Figures shows $\gamma_{max}$ on the left and $k_z^c$ on the right as a function of $Rm$ for different values of $Re$ mentioned in the legend. The curves correspond to the nonhelical forcing case.}
\label{Fig:gamma_kzcnonhel}
\end{figure}

\begin{figure}
\begin{center}
\includegraphics[scale=0.13]{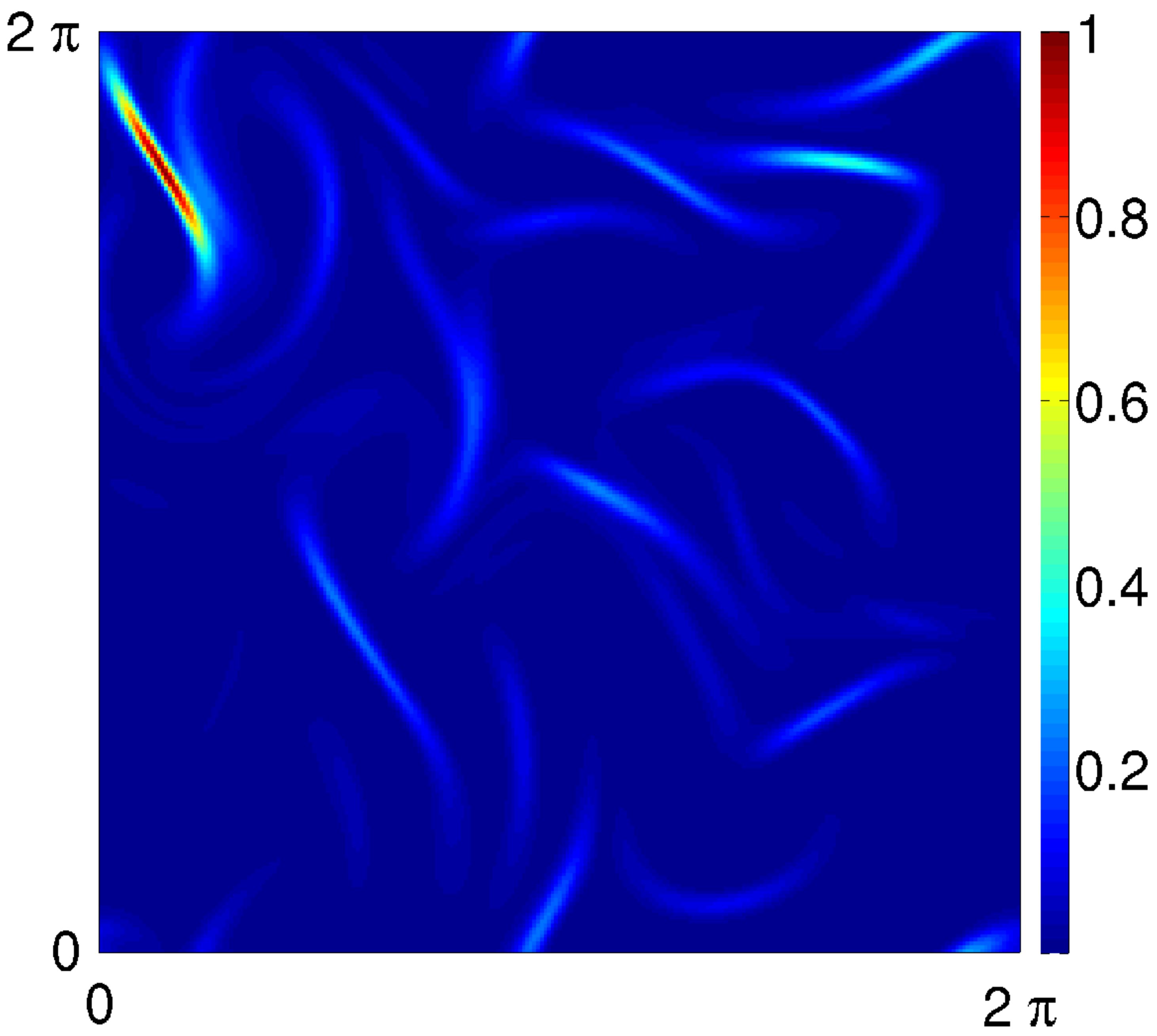} \hspace{2mm}
\includegraphics[scale=0.13]{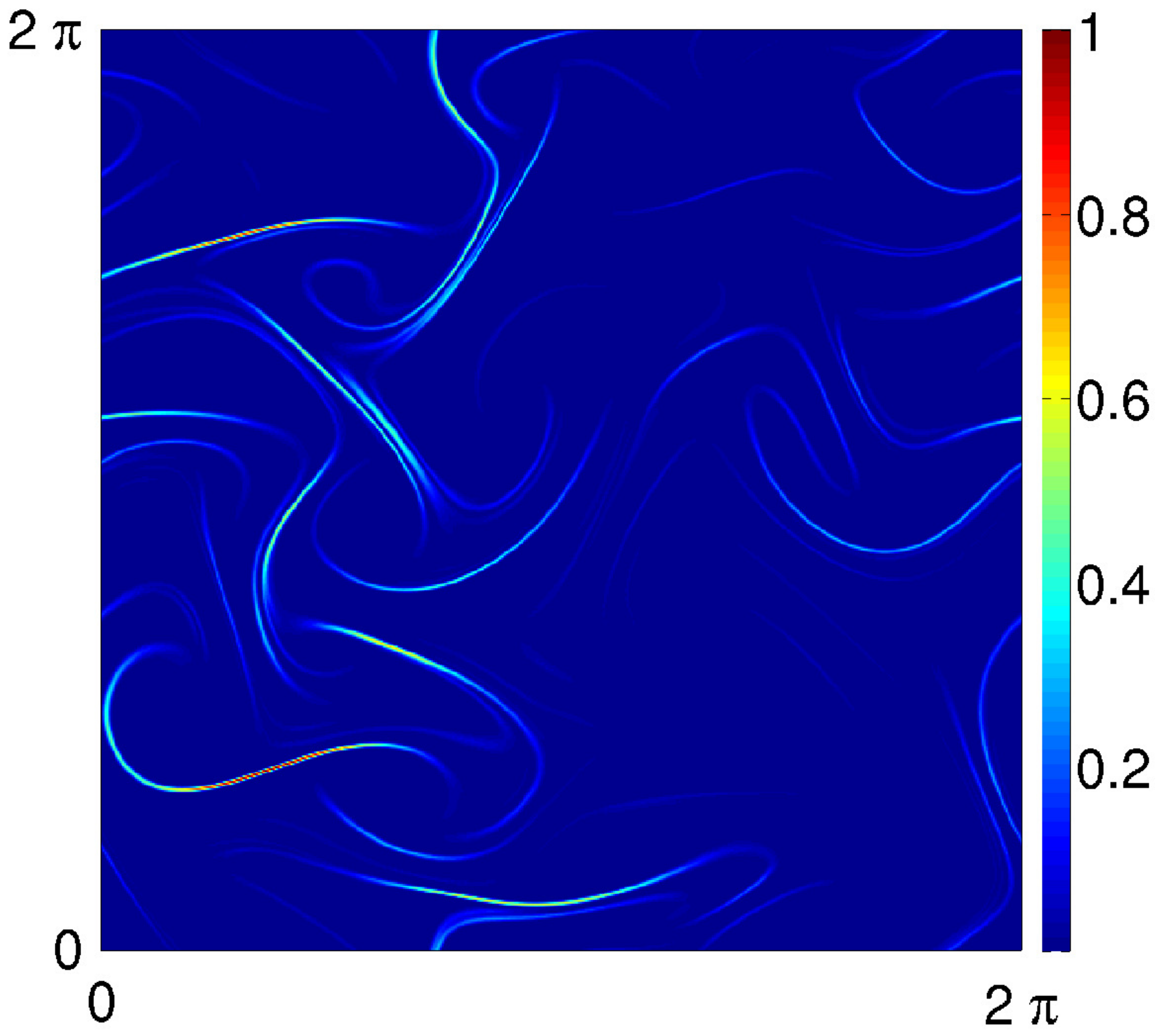} \hspace{2mm}
\includegraphics[scale=0.13]{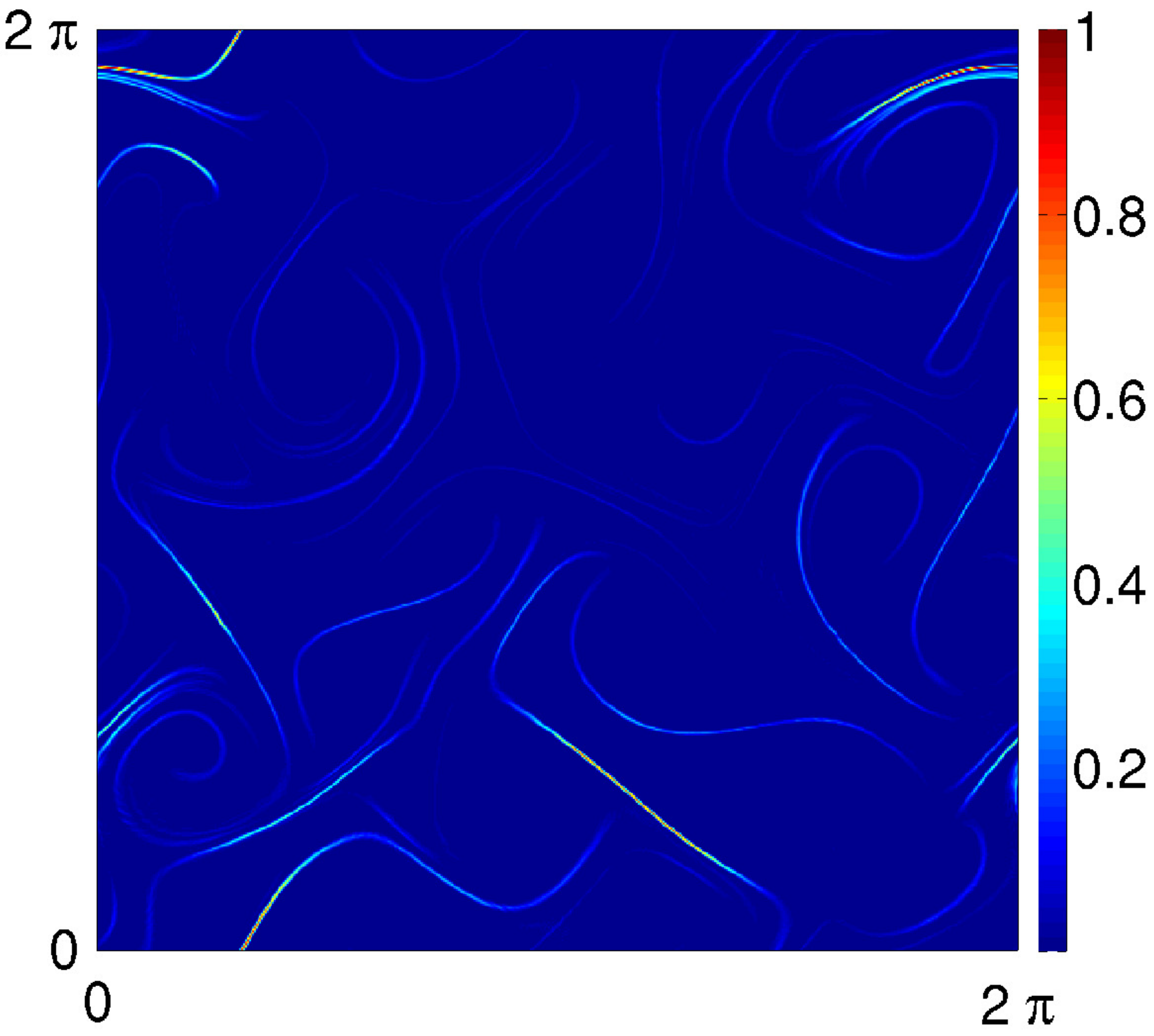}
\end{center}
\caption{Contour of the magnetic energy $B_{_{2D}}$ for different values of $Rm$, from left to right we have, $Rm \approx 32$, $Rm \approx 1030$, $Rm \approx 2060$ with the $Re \approx 32$ for all the three contours. The figures correspond to the nonhelical forcing case.}
\label{Fig:turb_cont}
\end{figure}

Figure \ref{Fig:gamma_kzcnonhel} shows $\gamma_{max}$ and $k_z^c$ as a function of $Re, Rm$. The dynamo instability starts at a $Rm \approx 10$ which is the critical magnetic Reynolds number for this type of forcing. Unlike the helical case the maximum growth rate $\gamma_{max}$ increases slowly with $Rm$ and a clear asymptote has not yet been reached. $Re$ does not seem to affect the behaviour of the $\gamma_{max}$ curve indicating that the most unstable modes are not affected by the smallest viscous scales. 
The scaling of $k_z^c \sim Rm^{1/2}$ in the large $Rm$ limit is observed with a prefactor that decreases as $Re$ is increased similar to the helical case. The magnetic field generated in the small scales is spatially concentrated in thin filamentary structures. Figure \ref{Fig:turb_cont} shows the contours of magnetic energy in the plane - $|B_{_{2D}}|^2 = |b_x|^2 + |b_y|^2$ for increasing values of the magnetic Reynolds number $Rm$. 
These structures become thiner 
as we increase $Rm$ with the thickness scaling like $Rm^{-1/2}$. 
This gives a physical interpretation for the scaling $k_z^c \sim Rm^{1/2}$ seen in figures \ref{Fig:gamma_kzchel}, \ref{Fig:gamma_kzcnonhel} in terms of $H$:
these filaments should be thinner than the box height $H$ for the dynamo instability to take place.
\section{Critical magnetic Reynolds number $Rm_c$} \label{Section:Six}

\subsection{Finite layer thickness \label{Sec:Finitedomain}}
In general the onset of the dynamo instability depends on the domain size since it determines the available wavemodes.
For a given height $H$ the allowed wavenumbers satisfy $k_z \geq 2 \pi/H \equiv k_z^H$. 
We thus define a critical magnetic Reynolds number $Rm_c$ based on $H$ as,
\begin{eqnarray}	
Rm_c^H  \, \left( Re, k_z^H \right) \; = \; \max \Big\{ Rm \;\; \text{s.t.} \;\; \gamma \leq 0 \;\;\; \forall k_z>k_z^H \Big\}. 
\end{eqnarray}
The dynamo instability then only exists for $Rm > Rm_c^H$. The value $Rm_c^H$ can be calculated from the figures \ref{Fig:gamma_kzchel},\ref{Fig:gamma_kzcnonhel} imposing that
the marginal $k_z$ for dynamo equals the minimum allowed wavenumber $k_z^c (Re,Rm) = k_z^H$. For the helical case in the small $Rm$ limit we get the relation $Rm_c^H \propto \sqrt{k_z^H}$ based on the $\alpha$-effect. 
Thus for large $H$ a small $Rm$ is sufficient for dynamo instability $Rm_c^H \propto (H)^{-1/2}$ with the proportionality coefficient being independent of $Re$.

\begin{figure}
\begin{center}
\includegraphics[scale=0.13]{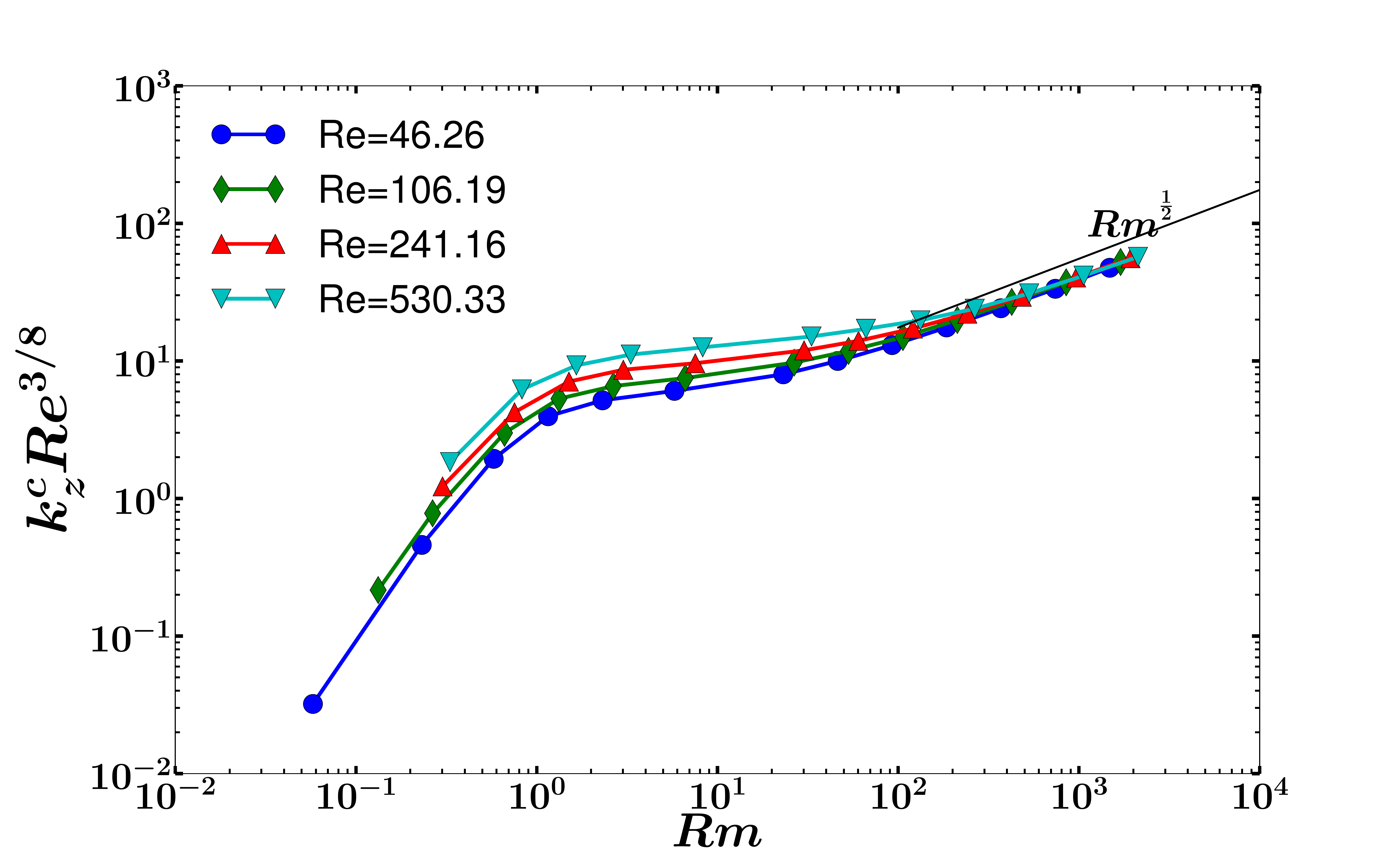}
\includegraphics[scale=0.13]{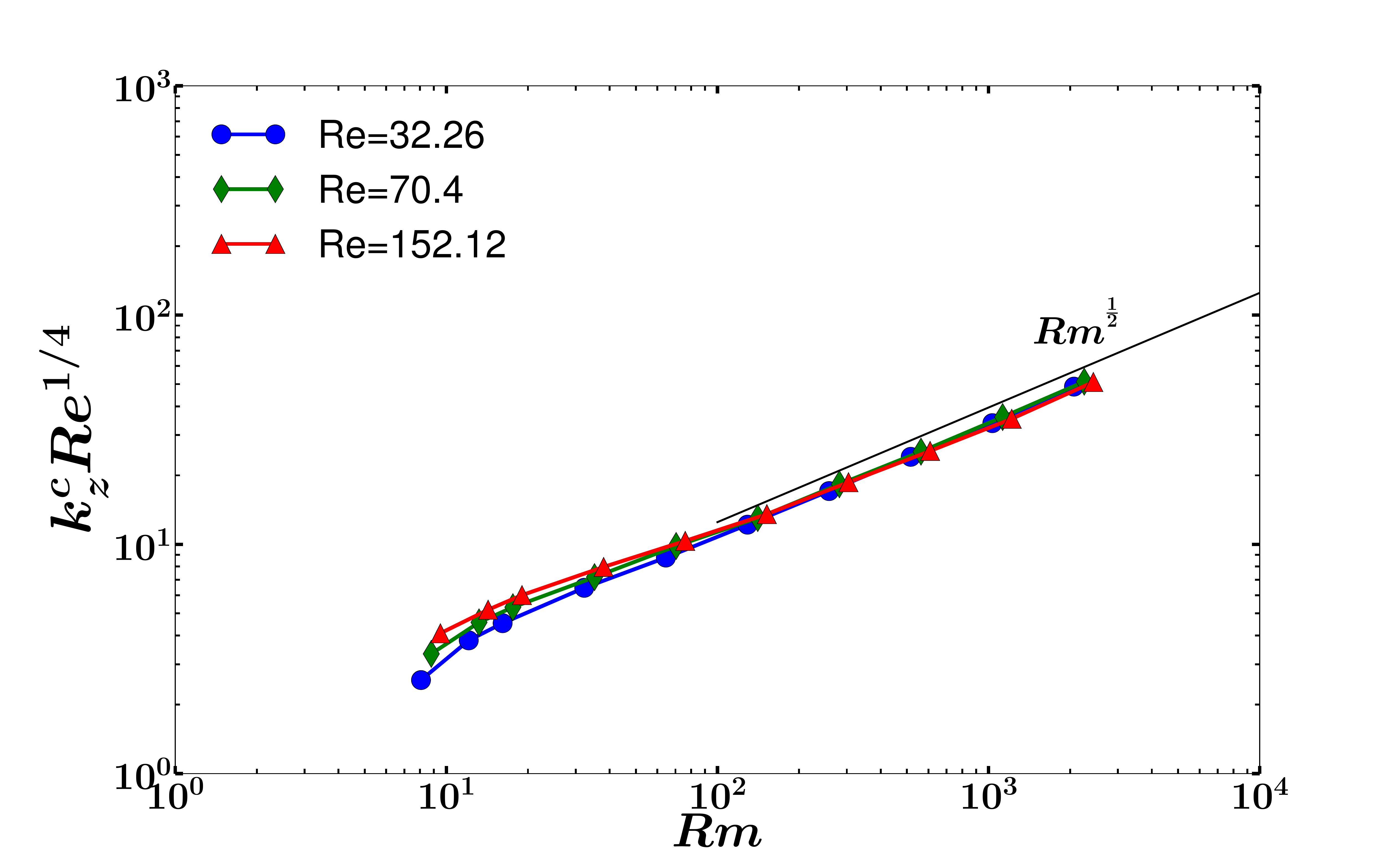}
\end{center}
\caption{Figure shows $k_z^c \, Re^{\zeta}$ as a function of $Rm$ for different values of $Re$ shown in the legend, for the helical forcing case shown on left and the nonhelical forcing case shown on the right.}
\label{Fig:finitebox_Rmc}
\end{figure}

The behaviour of $Rm^H_c$ for thin layers ($k_z^H\gg k_f$) depends on $Re$ for both the forcing cases considered. 
In order mesure this dependence on $Re$, we rescale $k_z^c$ with $Re$ and replot it as a function of $Rm$.
Figure \ref{Fig:finitebox_Rmc} shows the rescaled cut-off wavenumber $k_z^c \, Re^{\zeta}$ for the two different types of forcing studied.
Here $\zeta$ is an exponent used to collapse the data at large $k_z$. For the helical forcing we find a best fit of $\zeta = 0.37\dots \approx 3/8$ and for the nonhelical forcing we find a best fit of $\zeta = 0.25\dots \approx 1/4$. This implies that the critical magnetic Reynolds number scales like $Rm_c^H \propto \, Re^{2 \zeta}\,\sqrt{k_z^H} $. This is unlike the three dimensional dynamos where $Rm_c$ is found to reach a constant value in the large $Re$ limit. However, given that $\zeta<1/2$, in the limit of large $Re$, $Rm_c^H \ll Re$ thus like three dimensional turbulence dynamo can be achieved for any Prandtl number
$Pm=Rm/Re$ provided $R_m$ is large enough. Whether this behaviour persists for very large $Re$ remains to be seen.
\subsection{Infinite layer thickness }
\begin{figure}
\begin{center}
\includegraphics[scale=0.2]{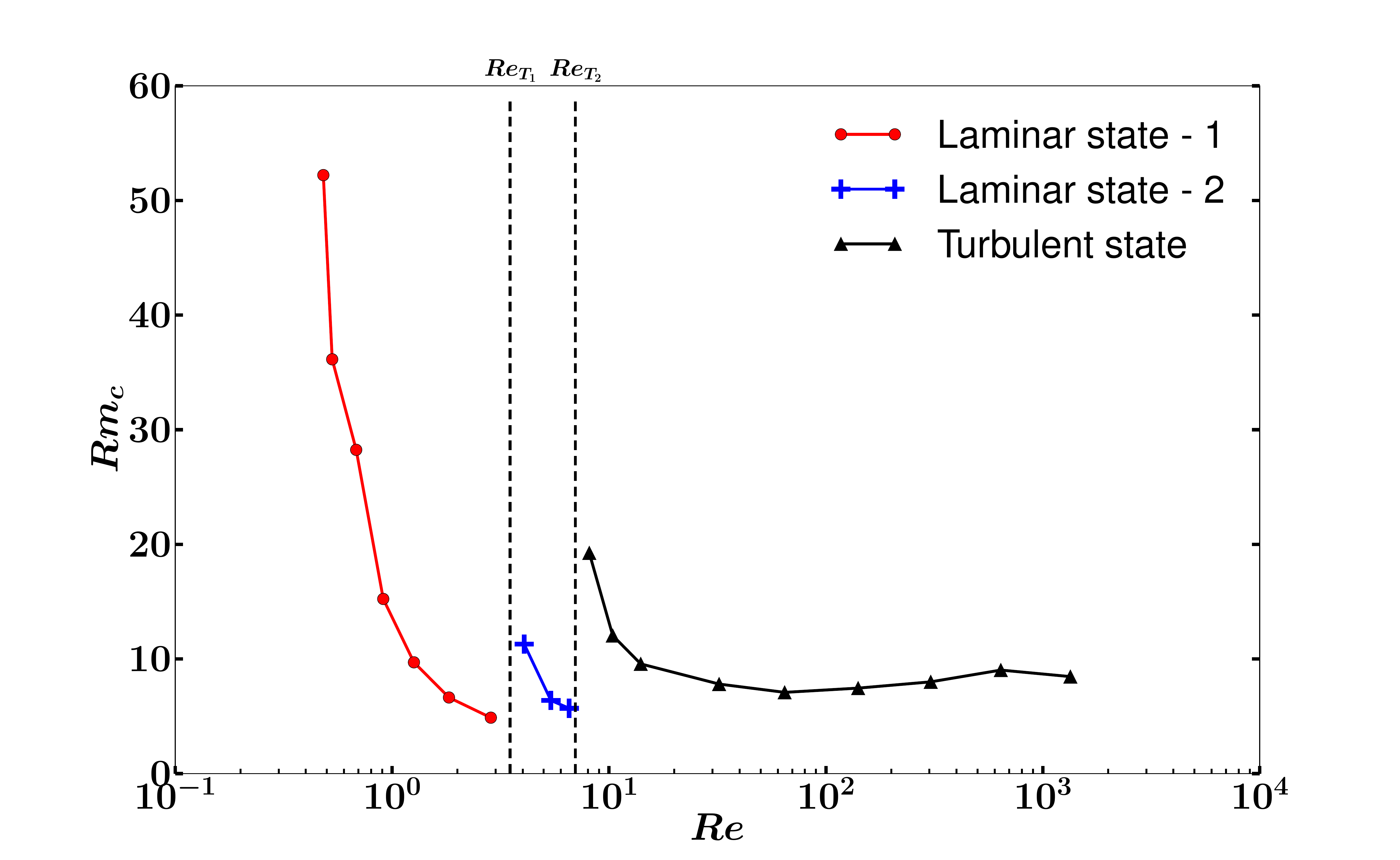}
\end{center}
\caption{Plot shows the critical magnetic Reynolds number $Rm_c$ as a function of the fluid Reynolds number $Re$. Two vertical dotted lines denote the two transition Reynolds numbers $Re_{T_1}, Re_{T_2}$. The curves correspond to the nonhelical forcing case.}
\label{Fig:Rmc}
\end{figure}

\begin{figure}
\begin{center}
\includegraphics[scale=0.15]{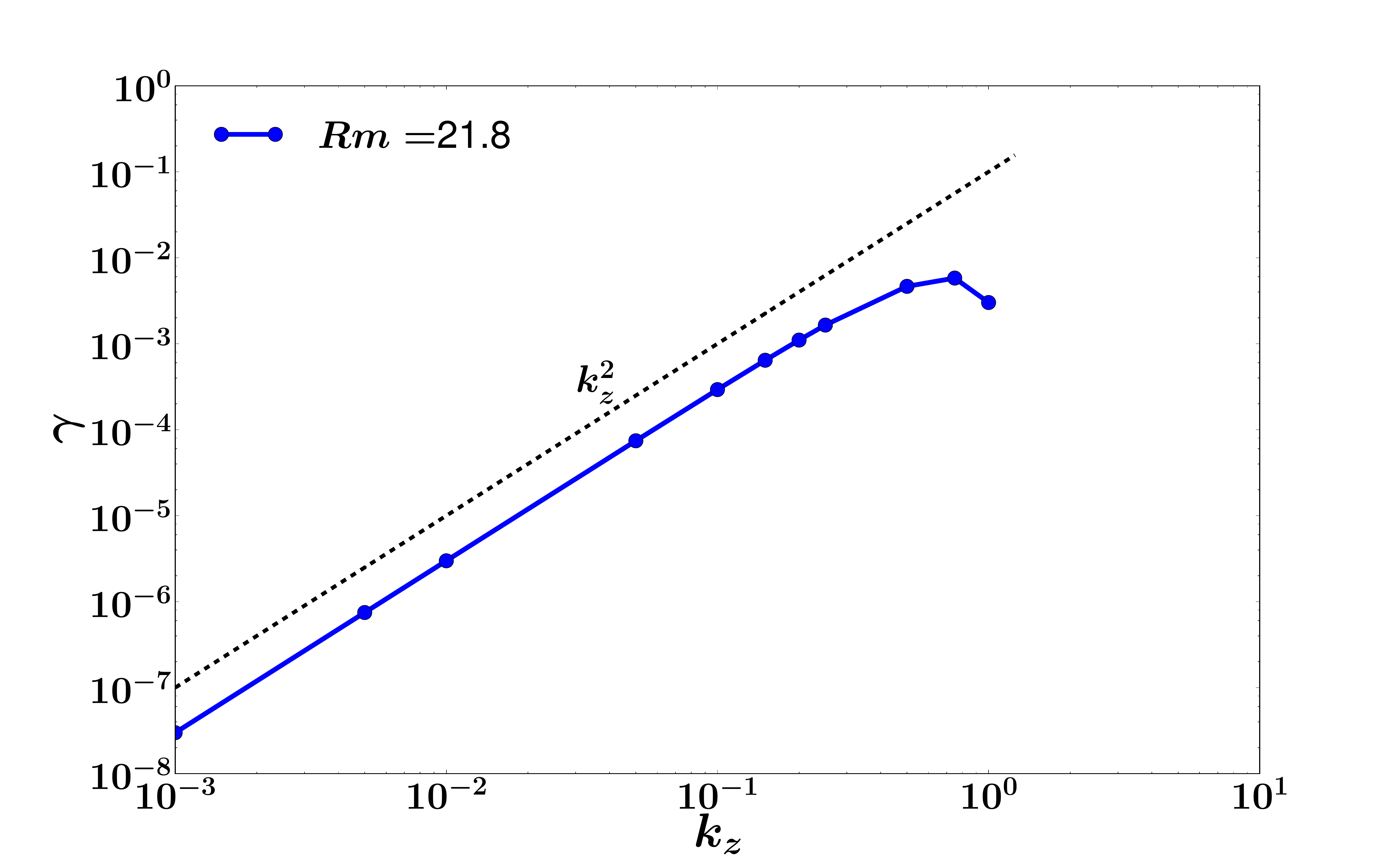}
\end{center}
\caption{Plot shows the growth rate $\gamma$ as a function of $k_z$ for a Reynolds number $Re=0.91 < Re_{T_1}$ is shown along with the dotted line with the scaling $k_z^2$. The curve correspond to the nonhelical forcing case.}
\label{Fig:betaeffect}
\end{figure}

\begin{figure}
\begin{center}
\includegraphics[scale=0.15]{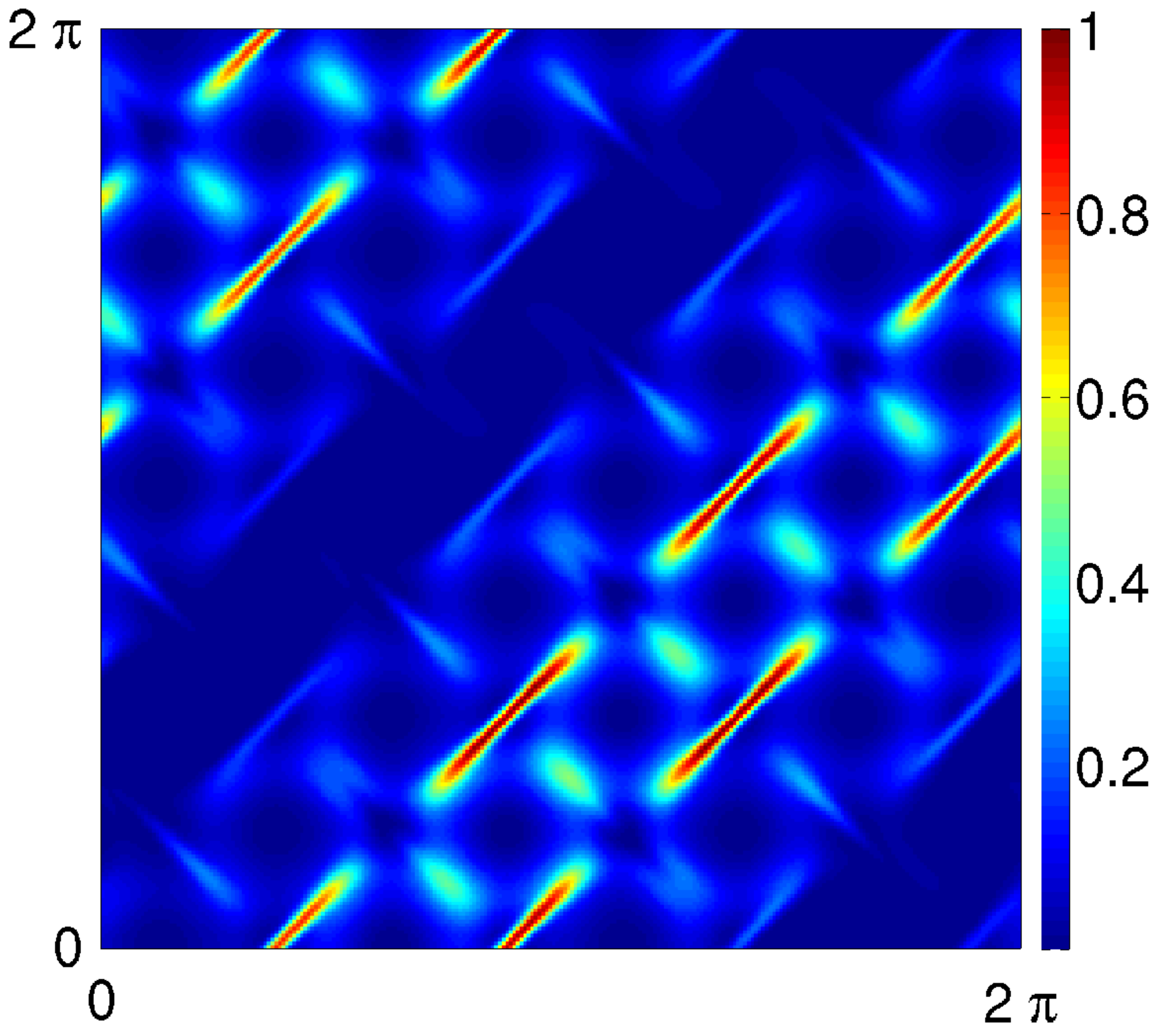} \hspace{10mm}
\includegraphics[scale=0.15]{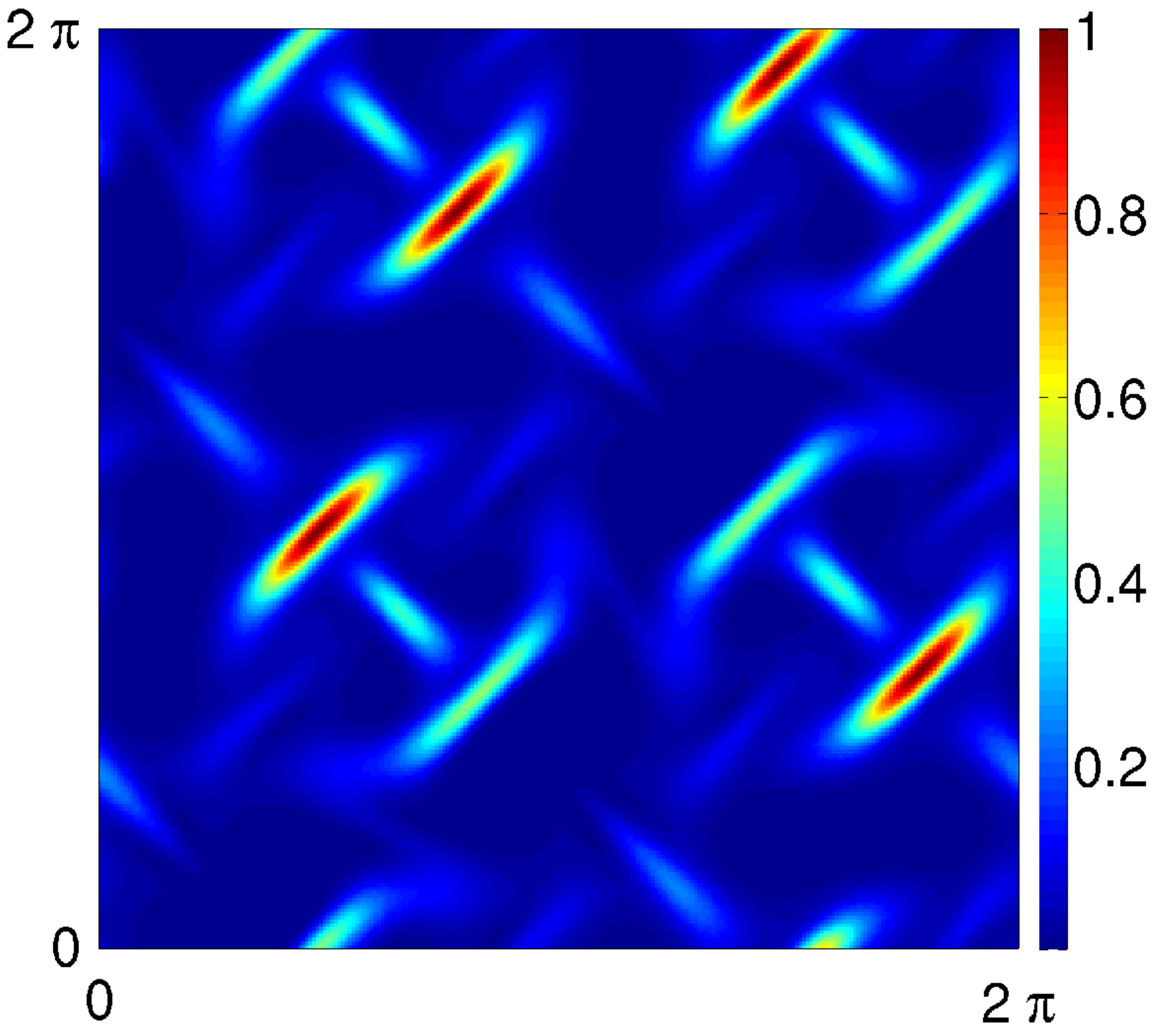}
\end{center}
\caption{Contour of the magnetic energy $B_{_{2D}}$ for the two different laminar flows at two different $Re$ - Left $Re_{T_1}<Re \approx 5.4<Re_{T_2}$, Right $Re \approx 0.53 < Re_{T_1}$. The contours correspond to the nonhelical forcing case. }
\label{Fig:lam_cont}
\end{figure}

As seen in figure \ref{Fig:gammavskzhel}, in helical flows due to the $\alpha$-effect for any $Rm$ there always exists $k_z$ small enough such that the modes are dynamo unstable. 
Thus for a layer that is infinitely thick, a helical flow does not have a critical magnetic Reynolds number 
since unstable modes exist even for $Rm\to 0$. For the nonhelical case however there is a critical $Rm$ for the dynamo instability as can be seen in figures \ref{Fig:gammavskznonhel}, \ref{Fig:gamma_kzcnonhel}. Below this $Rm_c$ for any mode $k_z$ there is no dynamo instability. 
Thus the critical magnetic Reynolds number $Rm_c$ in the infinite domain is defined as, 
\begin{eqnarray}	
Rm_c  \, \left( Re \right) \; = \; \max_{} \Big\{ Rm \;\; \text{s.t.} \;\; \gamma \leq 0 \;\;\; \forall k_z \Big\} = \lim_{H \rightarrow \infty} Rm_c^H. 
\end{eqnarray}
Note that in practice we do not need an infinitely thick layer to capture the onset of the instability. The height $H$ however needs to be sufficiently 
large so that it allows the first unstable mode $k_z\simeq 1$ (as can be seen in figure \ref{Fig:gammavskznonhel}) to be present. 
The dependence of $Rm_c$ as a function of $Re$ can be seen in the figure \ref{Fig:Rmc}. Three different regimes corresponding to different flow behaviours are identified and are separated by vertical dotted lines in the figure denoting the critical Reynolds numbers $Re_{T_1}, Re_{T_2}$. The curve for $Re>Re_{T_2}$ corresponds to the turbulent regime at large $Re$ and the curves in $Re<Re_{T_1}$, $Re_{T_1}<Re<Re_{T_2}$ correspond to two different laminar flows. Here $Re_{T_2}$ is the Reynolds number at which the flow transitions between a turbulent state and a laminar state. While $Re_{T_1}$ is the Reynolds number at which the flow transitions between two different laminar time independent flows. In the limit of large $Re$ we see that the value of $Rm_c$ saturates as is observed in $3D$ turbulent flows \cite{ponty2005numerical, iskakov2007numerical, mininni2007inverse}. Across the transition Reynolds numbers $Re_{T_2}$ and $Re_{T_1}$, the $Rm_c$ curves have discontinuous behaviour because the flow transitions from one state to the other subcritically.
In these laminar states we find that the growth rate $\gamma$ scales as $k_z^2$ for very small $k_z$ as shown in figure \ref{Fig:betaeffect} for a $Re=0.91<Re_{T_1}$ in the laminar regime. This scaling indicates that the dynamo action can be explained by the $\beta$-effect, also known in literature as the negative magnetic diffusivity effect, (see \cite{lanotte1999large}). The $\beta$-effect is a mean-field effect and the magnetic field is amplified also at the large scales. Figure \ref{Fig:lam_cont} shows the contour of the $|B_{_{2D}}|^2 = |b_x|^2 + |b_y|^2$ which is the energy of the magnetic field in the $x-y$ plane. Two different Reynolds number are shown, on the left $Re_{T_1}<Re=5.4<Re_{T_2}$ and on the right $Re=0.53<Re_{T_1}$ corresponding to the two different laminar states. Both the plots show large scale modulations in the magnetic energy at scales close to the box size. 


\section{Dependence on $k_f \, L$} \label{Section:Seven}

\begin{figure}
\begin{center}
\includegraphics[scale=0.13]{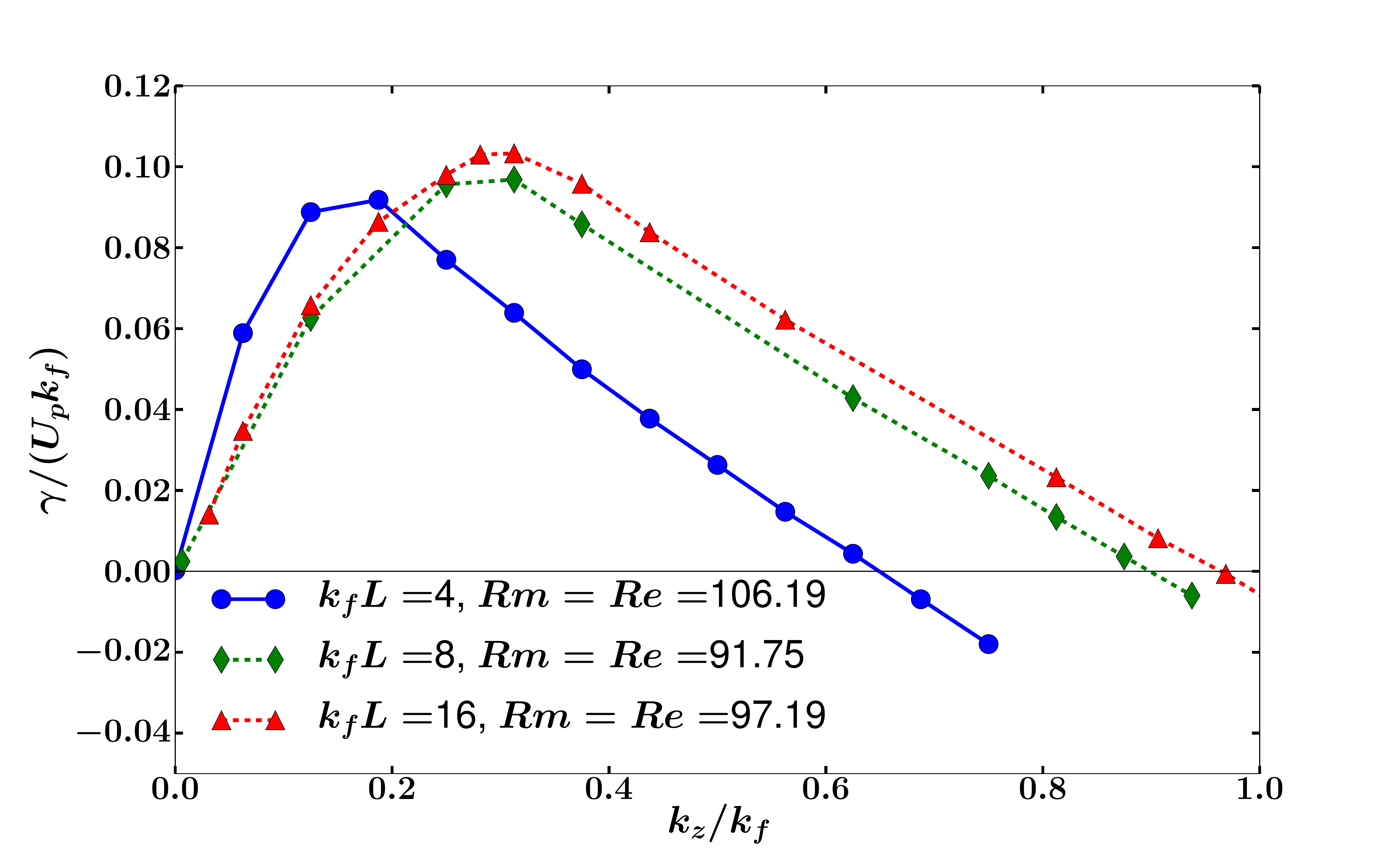}
\includegraphics[scale=0.13]{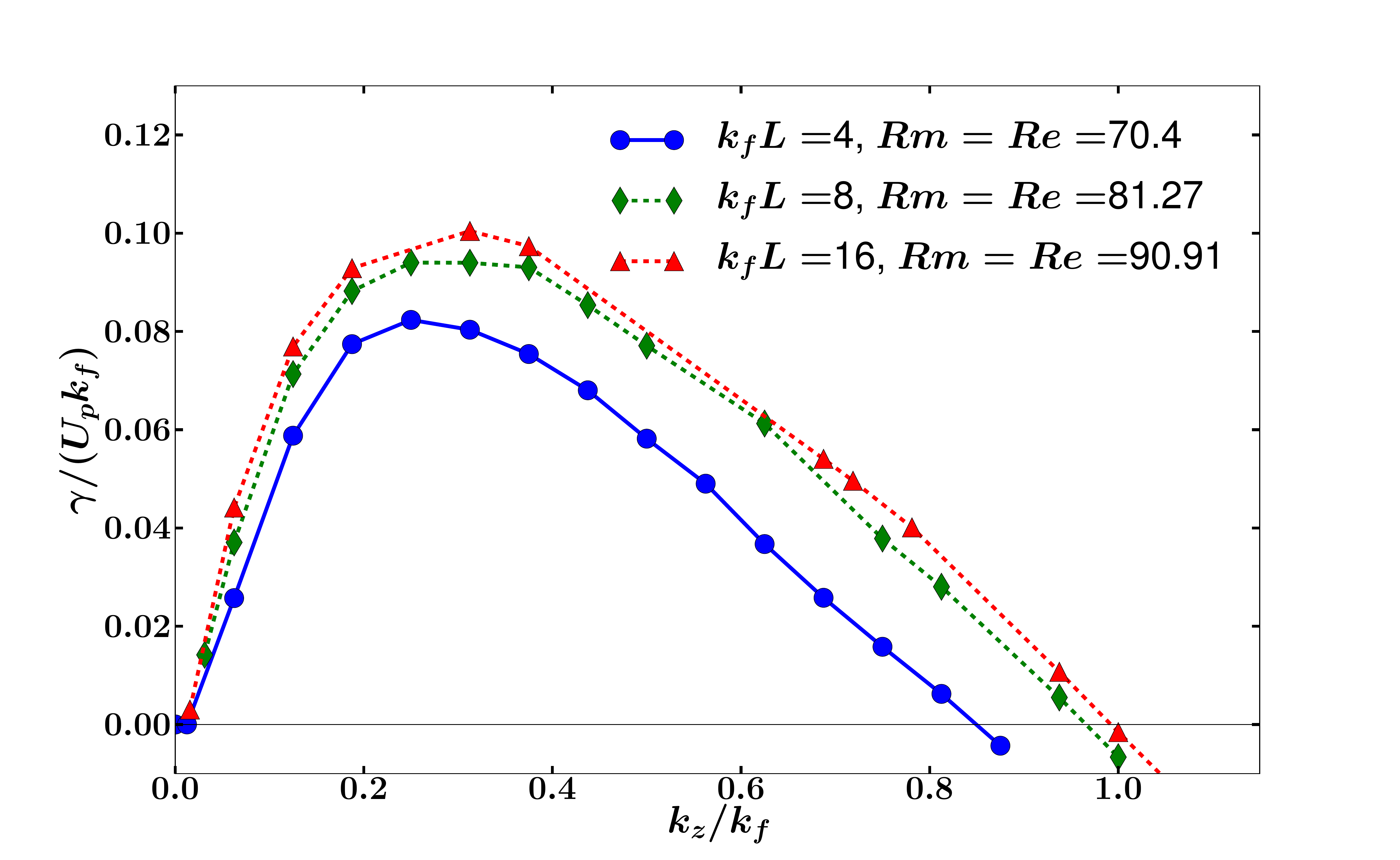}
\end{center}
\caption{Figure shows $\gamma/(U_p k_f)$ as a function of $k_z/k_f$ for different values of $k_f$ for helical forcing shown on left and nonhelical forcing shown on the right. The kinetic Reynolds number and the magnetic Reynolds number are mentioned in the legends.}
\label{Fig:gamma_kz_highkf}
\end{figure}

In this section we extend our study to flows with higher values of $k_f \, L$.
The linear damping coefficient is adjusted for each value of  $k_f \, L$ so that maximum inertial range for the inverse cascade is obtained without forming condensates.
As we increase $k_f \, L$ the large scale inverse cascade becomes more important. Depending on the forcing used and the scale seperation
the relative amplitude of $u_{_{2D}}$ and $u_z$ change as we change $k_f \, L$. 
In order thus to have a fair comparison between the different dynamos we normalize the growth rates based 
on the results of the Ponomarenko dynamo \citep{ponomarenko1973theory}, where the growth rate is proportional to the product of the vertical velocity $\bf u_z$ 
and the planar velocity $\bf u_{_{2D}}$ divided by the total rms value.
Thus we define a velocity scale, 
$U_p = (\left \langle |{\bf u_{_{2D}}}|^2 \right\rangle^{1/2} \, \left\langle u_z^2 \right\rangle^{1/2})/(\left\langle |{\bf u_{_{2D}}}| \right\rangle^2 + \left\langle |u_z| \right\rangle^2)^{1/2}$ 
with which we normalize the growth rate. Figure \ref{Fig:gamma_kz_highkf} shows normalized growth rate $\gamma/(U_p k_f)$ as a function of normalized modes $k_z/k_f$ for both the helical and nonhelical forcing as we increase $k_f \, L$ for similar values of $Re, Rm$. Since $k_f$ is increased the growth rate $\gamma$ and the number of unstable $k_z$ modes increase. This behaviour is similar for both the helical and nonhelical forcing case. The normalized curves seems to follow similar trend for both the forcing cases considered here. 
At relative large $Rm$ and as the scale separation is increased the most unstable wave number appears to be close to
the forcing wavenumber $k_z^{max} \approx \frac{1}{3} k_f$ in both helical and nonhelical forcing cases.
This implies that the most unstable modes have similar length scale with forcing and not with the box size.

\begin{figure}
\begin{center}
\includegraphics[scale=0.13]{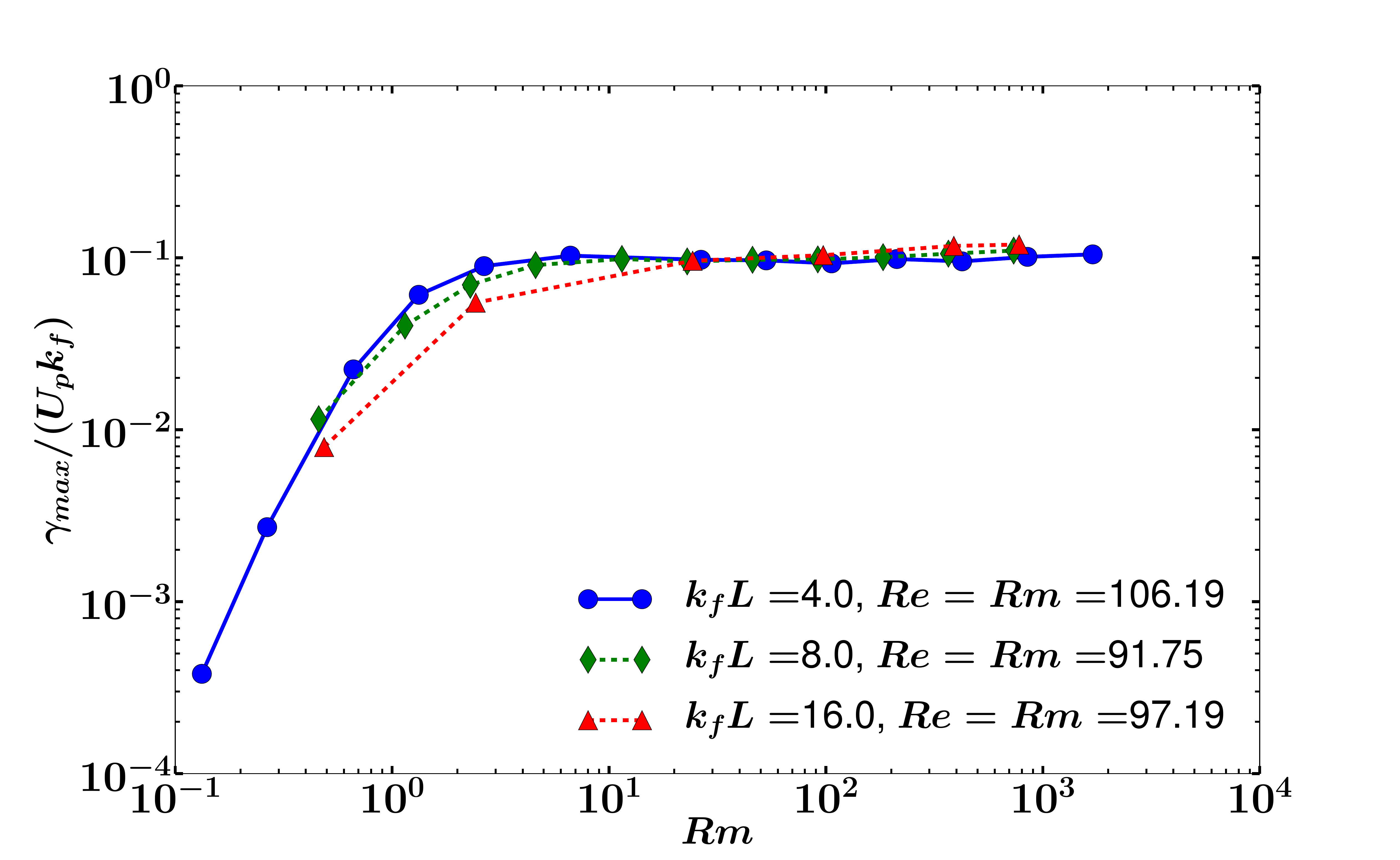}
\includegraphics[scale=0.13]{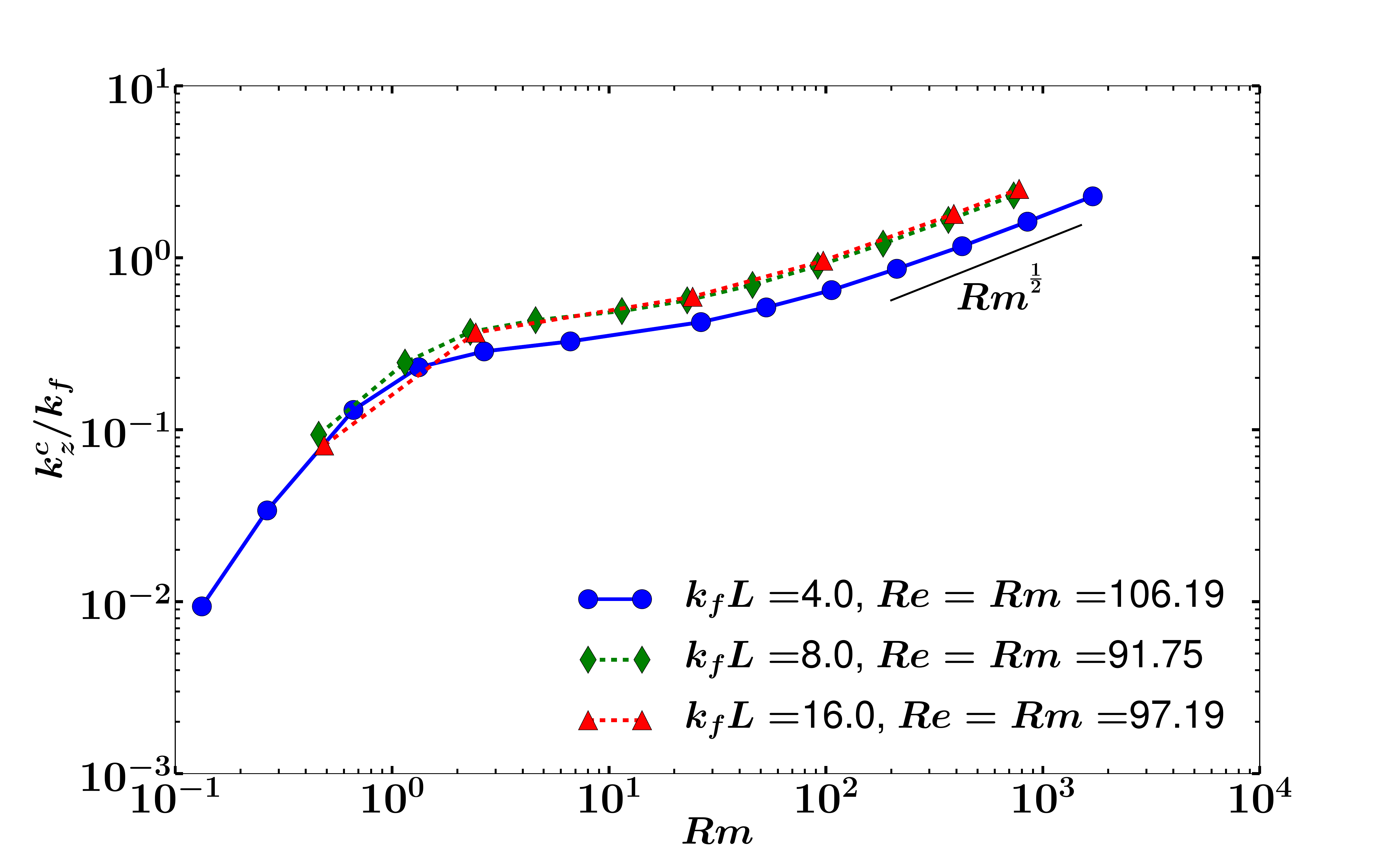}
\includegraphics[scale=0.13]{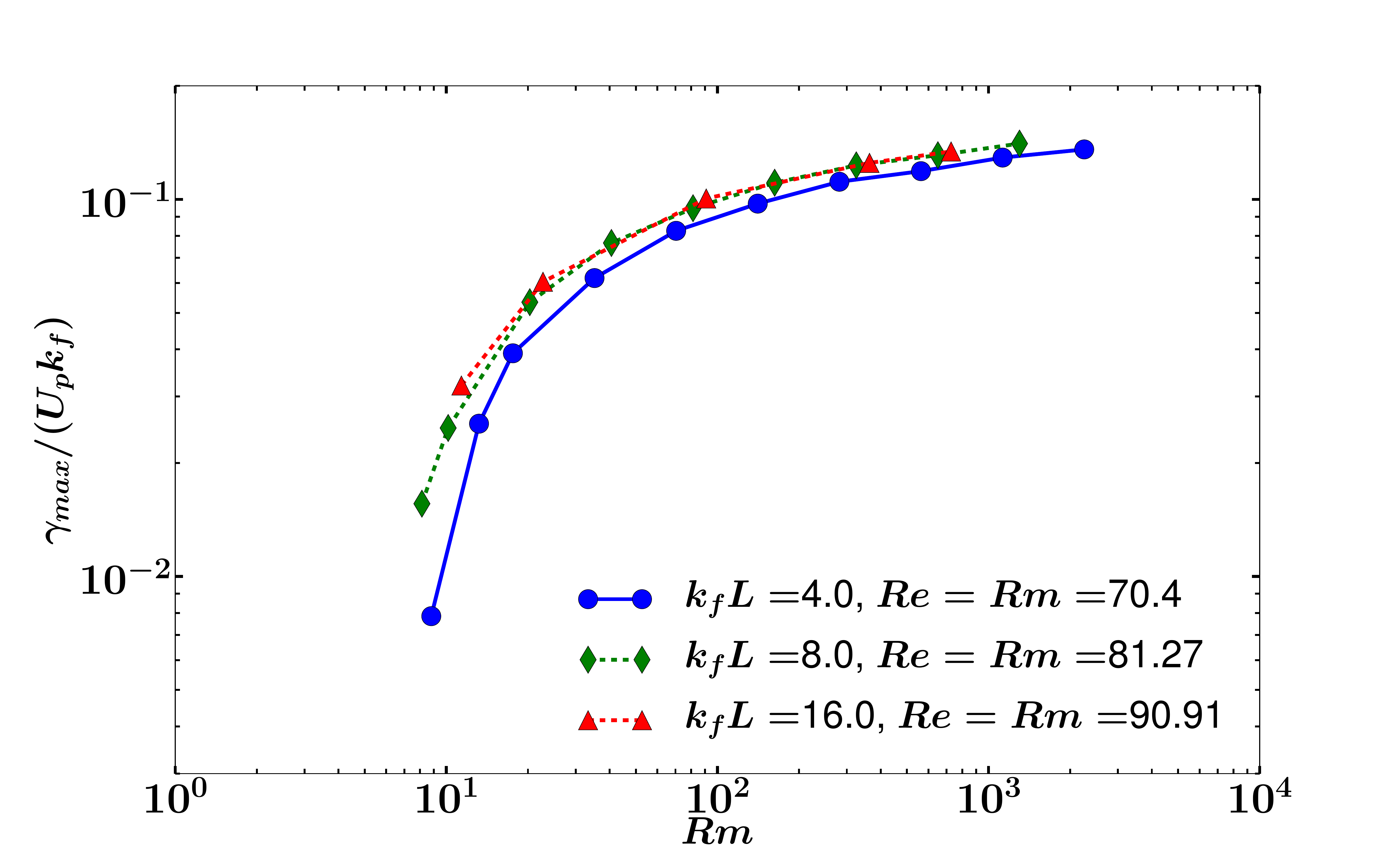}
\includegraphics[scale=0.13]{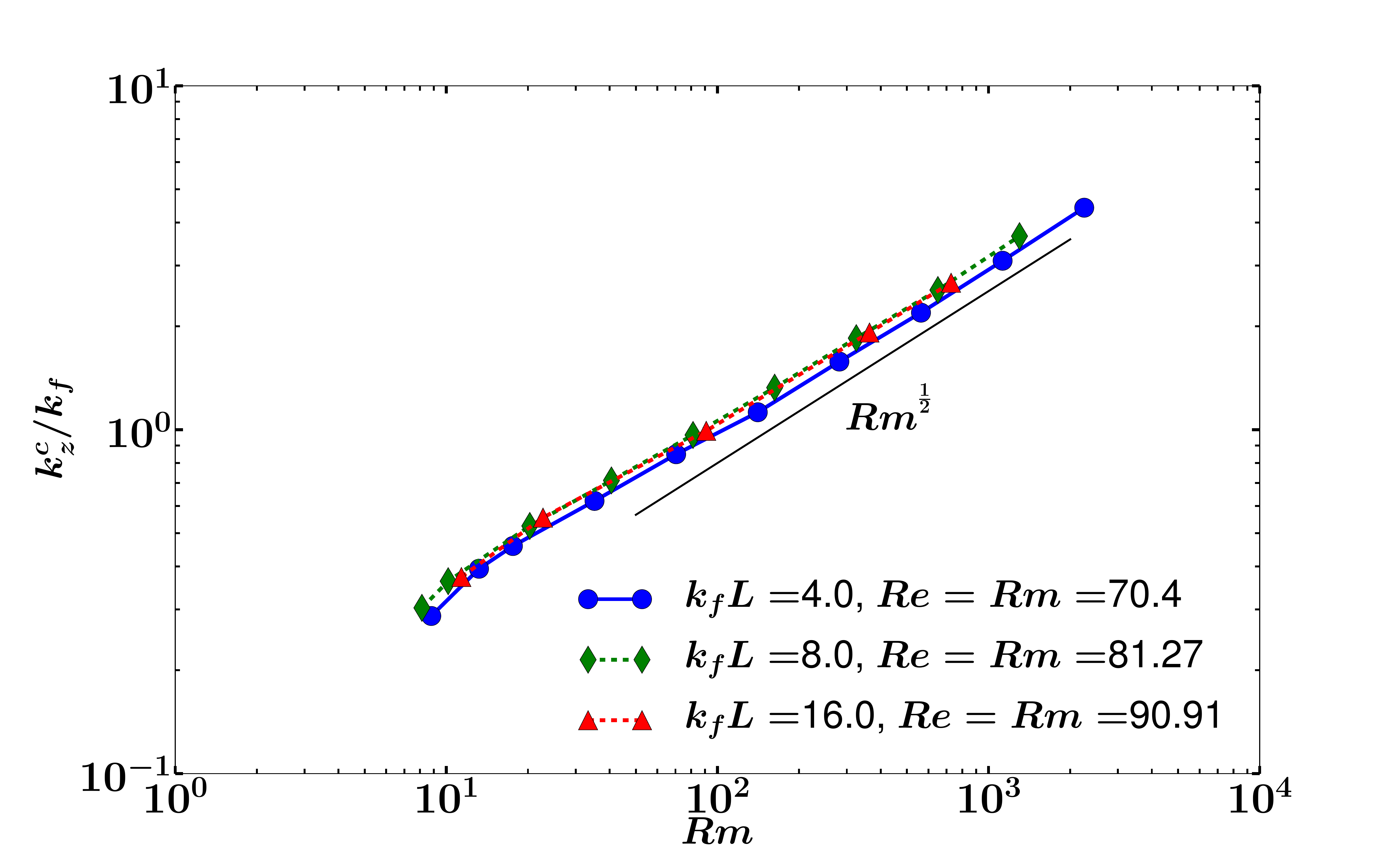}
\end{center}
\caption{Plots of normalized growth rate - $\gamma_{\max}/(U_p k_f)$ on the left and $k_z^c/k_f$ on the right for 1. Top - helical forcing and 2. Bottom - nonhelical forcing as a function of $Rm$ for different $k_f \, L$ mentioned in the legends.}
\label{Fig:highkf_quants}
\end{figure}

The normalized maximum growth rate $\gamma_{max}/(U_p k_f)$ and the normalized cut-off wavenumber $k_z^c/k_f$ for both helical and nonhelical forcing are shown in figure \ref{Fig:highkf_quants}. As can be seen from the figures the normalized quantities follow similar trends to $k_f \, L = 4$ with weak (or no) dependence on the box size $L$. 
Hence the inverse cascade does not seem to affect the dynamo instability, as is expected since the mechanisms of small scale dynamo effect and the $\alpha$-dynamo are mostly governed by the forcing scale or scales smaller than the forcing scale where the strongest shear exists.
%

\section{Conclusions} \label{Section:Eight}
The dynamo instability in the $2.5D$ configuration is studied for a wide range of control parameters. 
This allowed us to test certain limits that are still not attainable in three dimensional simulations.

For helical flows we were able to test the alpha dynamo predictions for the behaviour of the large scales ($k_z\ll k_f$) both for small and large values of $Rm,Re$.
The analytical predictions of mean field theories for small values of $Rm$ were verified. For large values of $Rm$ the growth rates were also shown to be in agreement with a turbulent 
alpha dynamo (calculated numerically from equations \ref{Eqn:alphacal}, \ref{Eqn:alphaeqn}), and the isotropic $\alpha$ was shown to asymptote to a value independent of $Re$ and $Rm$.
Nonetheless, at large $Rm$ the large scale modes were not the most unstable ones. At sufficiently large $Rm$ the fastest growing mode was always found to have $k_z$ 
close to the forcing wavenumber. Thus in a three dimensional simulation with random initial conditions for the magnetic field, it is the scales close
to the forcing that would be observed in the linear stage of the dynamo.
This of course does not imply that the large scale instability does not play a role in the saturated stage of the dynamo 
and the formation of large scale magnetic fields at high $Rm$. To resolve this issue however a nonlinear
formalism for the alpha dynamo would be required.

The non-helical forcing was also shown to result in dynamo instability above a value of the magnetic Reynolds number 
with similar behaviour in the small scales $k_z \gtrsim k_f$ as the helical dynamo. The critical value of the magnetic Reynolds number
for a thin layer of height $H$ was shown to scale like $Rm_c^H \propto Re^{2\zeta}/\sqrt{H}$ with $\zeta \simeq 1/4$ for nonhelical flows and
$\zeta \simeq 3/8$ for helical flows, implying that there is a dependence of $Rm_c^H$ on $Re$ even at large values of $Re$. At infinite layer thickness $H$ the helical flow always resulted in to dynamo (ie $Rm_c=0$). 
On the other hand the non-helical flow $Rm_c$ was reaching asymptotically a finite value in the limit $Re\to \infty$.

The investigated dynamo flows were motivated by rotating flows that tend to become two dimensional at sufficiently large rotation rates.
In nature rotating flows are never fully two-dimensionalized. Even in fast rotating flows large two-dimensional motions co-exist 
with three dimensional perturbations either in the form of turbulent eddies or travelling inertial waves. The resulting dynamo then 
is in general the result of a combination these effects. However, due to the fast decorrelation time of eddies and inertial waves
that has a suppressing effect for dynamo we expect that at fast rotating flows 2.5D flows could play the dominant effect
for dynamo. Rotation could thus also provide a mechanism to improve the dynamo experiments. Such an expectation can be verified by a study of the full three dimensional dynamo flow subject to fast rotation.

\acknowledgements

The authors would like to thank F. Petrelis, S. Fauve and B. Gallet for their very useful comments and fruitful discussions. 
The present work benefited from the computational support of the HPC resources of 
GENCI-TGCC-CURIE \& GENCI-CINES-JADE (Project No. x2014056421 \& No. x2015056421)
and MesoPSL financed by the Region Ile de France and the project EquipMeso (reference ANR-10-EQPX-29-01) where the
present numerical simulations have been performed.

%
%

\bibliographystyle{jfm}
\bibliography{refs}

\begin{thebibliography}{41}
\expandafter\ifx\csname natexlab\endcsname\relax\def\natexlab#1{#1}\fi
\def\au#1{#1} \def\ed#1{#1} \def\yr#1{#1}\def\at#1{#1}\def\jt#1{\textit{#1}}
  \def\bt#1{#1}\def\bvol#1{\textbf{#1}} \def\vol#1{#1} \def\pg#1{#1}
  \def\publ#1{#1}\def\arxiv#1{#1}\def\org#1{#1}\def\st#1{\textit{#1}}

\bibitem[Alexakis(2015)]{alexakis2015rotating}
{\sc \au{Alexakis, A.}} \yr{2015}  \at{Rotating taylor--green flow}.  \jt{J.
  Fluid Mech.}  \bvol{769},  \pg{46--78}.

\bibitem[{Bartello} {\em et~al.\/}(1994){Bartello}, {Metais} \&
  {Lesieur}]{Bartello1994}
{\sc \au{{Bartello}, P.}, \au{{Metais}, O.} \& \au{{Lesieur}, M.}} \yr{1994}
  \at{{Coherent structures in rotating three-dimensional turbulence}}.  \jt{J.
  Fluid Mech.}  \bvol{273},  \pg{1--29}.

\bibitem[Batchelor(1959)]{batchelor1959small}
{\sc \au{Batchelor, G.~K.}} \yr{1959}  \at{Small-scale variation of convected
  quantities like temperature in turbulent fluid part 1. general discussion and
  the case of small conductivity}.  \jt{J. Fluid Mech.}  \bvol{5}~(01),
  \pg{113--133}.

\bibitem[Boffetta(2007)]{boffetta2007energy}
{\sc \au{Boffetta, G.}} \yr{2007}  \at{Energy and enstrophy fluxes in the
  double cascade of two-dimensional turbulence}.  \jt{J. Fluid Mech.}
  \bvol{589},  \pg{253--260}.

\bibitem[{Campagne} {\em et~al.\/}(2014){Campagne}, {Gallet}, {Moisy} \&
  {Cortet}]{Campagne2014direct}
{\sc \au{{Campagne}, A.}, \au{{Gallet}, B.}, \au{{Moisy}, F.} \& \au{{Cortet},
  P.-P.}} \yr{2014}  \at{{Direct and inverse energy cascades in a forced
  rotating turbulence experiment}}.  \jt{Phys. Fluids}  \bvol{26}~(12),
  \pg{125112}.

\bibitem[{Chen} {\em et~al.\/}(2005){Chen}, {Chen}, {Eyink} \&
  {Holm}]{Chen2005}
{\sc \au{{Chen}, Q.}, \au{{Chen}, S.}, \au{{Eyink}, G.~L.} \& \au{{Holm},
  D.~D.}} \yr{2005}  \at{{Resonant interactions in rotating homogeneous
  three-dimensional turbulence}}.  \jt{J. Fluid Mech.}  \bvol{542},
  \pg{139--164}.

\bibitem[Childress(1969)]{childress1969class}
{\sc \au{Childress, S.}} \yr{1969}  \at{A class of solutions of the
  magnetohydrodynamic dynamo problem}.  \jt{The Application of Modern Physics
  to the Earth and Planetary Interiors}  \pg{pp. 629--648}.

\bibitem[Courvoisier {\em et~al.\/}(2006)Courvoisier, Hughes \&
  Tobias]{courvoisier2006alpha}
{\sc \au{Courvoisier, A.}, \au{Hughes, D.~W.} \& \au{Tobias, S.~M.}} \yr{2006}
  \at{$\alpha$ effect in a family of chaotic flows}.  \jt{Phys. Rev. Lett.}
  \bvol{96}~(3),  \pg{034503}.

\bibitem[Davidson(2014)]{davidson2014dynamics}
{\sc \au{Davidson, P.~A.}} \yr{2014}  \at{The dynamics and scaling laws of
  planetary dynamos driven by inertial waves}.  \jt{Geophys. J. Int.}
  \bvol{198}~(3),  \pg{1832--1847}.

\bibitem[Deusebio {\em et~al.\/}(2014)Deusebio, Boffetta, Lindborg \&
  Musacchio]{deusebio2014dimensional}
{\sc \au{Deusebio, E.}, \au{Boffetta, G.}, \au{Lindborg, E.} \& \au{Musacchio,
  S.}} \yr{2014}  \at{Dimensional transition in rotating turbulence}.
  \jt{Phys. Rev. E}  \bvol{90}~(2),  \pg{023005}.

\bibitem[Gallet(2015)]{Gallet2015exact}
{\sc \au{Gallet, B.}} \yr{2015}  \at{Exact two-dimensionalization of rapidly
  rotating large-reynolds-number flows}.  \jt{J. Fluid Mech.}  \bvol{783},
  \pg{412--447}.

\bibitem[{Gallet} {\em et~al.\/}(2014){Gallet}, {Campagne}, {Cortet} \&
  {Moisy}]{Gallet2014}
{\sc \au{{Gallet}, B.}, \au{{Campagne}, A.}, \au{{Cortet}, P.-P.} \&
  \au{{Moisy}, F.}} \yr{2014}  \at{{Scale-dependent cyclone-anticyclone
  asymmetry in a forced rotating turbulence experiment}}.  \jt{Phys. Fluids}
  \bvol{26}~(3),  \pg{035108}.

\bibitem[Galloway \& Proctor(1992)]{galloway1992numerical}
{\sc \au{Galloway, D.~J.} \& \au{Proctor, M. R.~E.}} \yr{1992}  \at{Numerical
  calculations of fast dynamos in smooth velocity fields with realistic
  diffusion}.  \jt{Nature}  \bvol{356},  \pg{691 -- 693}.

\bibitem[Gilbert(2003)]{gilbert2003dynamo}
{\sc \au{Gilbert, A.~D.}} \yr{2003}  \at{Dynamo theory}.  \jt{Handbook of
  mathematical fluid dynamics}  \bvol{2},  \pg{355--441}.

\bibitem[Gomez {\em et~al.\/}(2005)Gomez, Mininni \& Dmitruk]{Gomez05}
{\sc \au{Gomez, D.~O.}, \au{Mininni, P.~D.} \& \au{Dmitruk, P.}} \yr{2005}
  \at{Parallel simulations in turbulent mhd}.  \jt{Phys. Scr.}  \bvol{T 116},
  \pg{123}.

\bibitem[{Hopfinger} \& {van Heijst}(1993)]{Hopfinger1993}
{\sc \au{{Hopfinger}, E.~J.} \& \au{{van Heijst}, G.~J.~F.}} \yr{1993}
  \at{{Vortices in rotating fluids}}.  \jt{Annu. Rev. Fluid Mech.}  \bvol{25},
  \pg{241--289}.

\bibitem[{Hossain}(1994)]{Hossain1994}
{\sc \au{{Hossain}, M.}} \yr{1994}  \at{{Reduction in the dimensionality of
  turbulence due to a strong rotation}}.  \jt{Phys. Fluids}  \bvol{6},
  \pg{1077--1080}.

\bibitem[Iskakov {\em et~al.\/}(2007)Iskakov, Schekochihin, Cowley, McWilliams
  \& Proctor]{iskakov2007numerical}
{\sc \au{Iskakov, A.~B.}, \au{Schekochihin, A.~A.}, \au{Cowley, S.~C.},
  \au{McWilliams, J.~C.} \& \au{Proctor, M. R.~E.}} \yr{2007}  \at{Numerical
  demonstration of fluctuation dynamo at low magnetic prandtl numbers}.
  \jt{Phys. Rev. Lett.}  \bvol{98}~(20),  \pg{208501}.

\bibitem[{Izakov}(2013)]{Izakov2013}
{\sc \au{{Izakov}, M.~N.}} \yr{2013}  \at{{Large-scale quasi-two-dimensional
  turbulence and a inverse spectral flux of energy in the atmosphere of
  Venus}}.  \jt{Solar Syst. Res.}  \bvol{47},  \pg{170--181}.

\bibitem[Lanotte {\em et~al.\/}(1999)Lanotte, Noullez, Vergassola \&
  Wirth]{lanotte1999large}
{\sc \au{Lanotte, A.}, \au{Noullez, A.}, \au{Vergassola, M.} \& \au{Wirth, A.}}
  \yr{1999}  \at{Large-scale dynamo produced by negative magnetic eddy
  diffusivities}.  \jt{Geophys. Astrophys. Fluid Dyn.}  \bvol{91}~(1-2),
  \pg{131--146}.

\bibitem[Mininni(2007)]{mininni2007inverse}
{\sc \au{Mininni, PD}} \yr{2007}  \at{Inverse cascades and $\alpha$ effect at a
  low magnetic prandtl number}.  \jt{Phys. Rev. E}  \bvol{76}~(2),
  \pg{026316}.

\bibitem[{Mininni} \& {Pouquet}(2010)]{Mininni2010}
{\sc \au{{Mininni}, P.~D.} \& \au{{Pouquet}, A.}} \yr{2010}  \at{{Rotating
  helical turbulence. I. Global evolution and spectral behavior}}.  \jt{Phys.
  Fluids}  \bvol{22}~(3),  \pg{035105}.

\bibitem[Pedlosky(1987)]{pedlosky2013geophysical}
{\sc \au{Pedlosky, J.}} \yr{1987} {\em Geophysical fluid dynamics\/}.
  \publ{New York, Springer}.

\bibitem[Ponomarenko(1973)]{ponomarenko1973theory}
{\sc \au{Ponomarenko, Yu~B}} \yr{1973}  \at{On the theory of the hydrodynamic
  dynamo}.  \jt{J. Appl. Mech. Tech. Phys}  \bvol{14},  \pg{775--779}.

\bibitem[Ponty {\em et~al.\/}(2005)Ponty, Mininni, Montgomery, Pinton, Politano
  \& Pouquet]{ponty2005numerical}
{\sc \au{Ponty, Y.}, \au{Mininni, P.~D.}, \au{Montgomery, D.~C.}, \au{Pinton,
  J-F.}, \au{Politano, H.} \& \au{Pouquet, A.}} \yr{2005}  \at{Numerical study
  of dynamo action at low magnetic prandtl numbers}.  \jt{Phys. Rev. Lett.}
  \bvol{94}~(16),  \pg{164502}.

\bibitem[{Proctor} \& {Gilbert}(1995)]{1995lspdP}
{\sc \au{{Proctor}, M.~R.~E.} \& \au{{Gilbert}, A.~D.}} \yr{1995} {\em
  {Lectures on Solar and Planetary Dynamos}\/}.

\bibitem[Roberts(1972)]{roberts1972dynamo}
{\sc \au{Roberts, G.~O.}} \yr{1972}  \at{Dynamo action of fluid motions with
  two-dimensional periodicity}.  \jt{Phil. Trans. R. Soc.}  \bvol{271}~(1216),
  \pg{411--454}.

\bibitem[{Scott}(2014)]{Scott2014}
{\sc \au{{Scott}, J.~F.}} \yr{2014}  \at{{Wave turbulence in a rotating
  channel}}.  \jt{J. Fluid Mech.}  \bvol{741},  \pg{316--349}.

\bibitem[{Sen} {\em et~al.\/}(2012){Sen}, {Mininni}, {Rosenberg} \&
  {Pouquet}]{Mininni2012}
{\sc \au{{Sen}, A.}, \au{{Mininni}, P.~D.}, \au{{Rosenberg}, D.} \&
  \au{{Pouquet}, A.}} \yr{2012}  \at{{Anisotropy and nonuniversality in scaling
  laws of the large-scale energy spectrum in rotating turbulence}}.  \jt{Phys.
  Rev. E}  \bvol{86}~(3),  \pg{036319}.

\bibitem[{Smith} \& {Waleffe}(1999)]{Smith1999}
{\sc \au{{Smith}, L.~M.} \& \au{{Waleffe}, F.}} \yr{1999}  \at{{Transfer of
  energy to two-dimensional large scales in forced, rotating three-dimensional
  turbulence}}.  \jt{Phys. Fluids}  \bvol{11},  \pg{1608--1622}.

\bibitem[Smith \& Tobias(2004)]{smith2004vortex}
{\sc \au{Smith, S. G.~L.} \& \au{Tobias, S.~M.}} \yr{2004}  \at{Vortex
  dynamos}.  \jt{J. Fluid Mech.}  \bvol{498},  \pg{1--21}.

\bibitem[{Staplehurst} {\em et~al.\/}(2008){Staplehurst}, {Davidson} \&
  {Dalziel}]{Staplehurst2008}
{\sc \au{{Staplehurst}, P.~J.}, \au{{Davidson}, P.~A.} \& \au{{Dalziel},
  S.~B.}} \yr{2008}  \at{{Structure formation in homogeneous freely decaying
  rotating turbulence}}.  \jt{J. Fluid Mech.}  \bvol{598},  \pg{81--105}.

\bibitem[{Sugihara} {\em et~al.\/}(2005){Sugihara}, {Migita} \&
  {Honji}]{Sugihara2005}
{\sc \au{{Sugihara}, Y.}, \au{{Migita}, M.} \& \au{{Honji}, H.}} \yr{2005}
  \at{{Orderly flow structures in grid-generated turbulence with background
  rotation}}.  \jt{Fluid Dyn. Res.}  \bvol{36},  \pg{23--34}.

\bibitem[{Thiele} \& {M{\"u}ller}(2009)]{Thiele2009}
{\sc \au{{Thiele}, M.} \& \au{{M{\"u}ller}, W.-C.}} \yr{2009}  \at{{Structure
  and decay of rotating homogeneous turbulence}}.  \jt{J. Fluid Mech.}
  \bvol{637},  \pg{425}.

\bibitem[Tobias \& Cattaneo(2008)]{tobias2008dynamo}
{\sc \au{Tobias, S.~M.} \& \au{Cattaneo, F.}} \yr{2008}  \at{Dynamo action in
  complex flows: the quick and the fast}.  \jt{J. Fluid Mech.}  \bvol{601},
  \pg{101--122}.

\bibitem[{van Bokhoven} {\em et~al.\/}(2009){van Bokhoven}, {Clercx}, {van
  Heijst} \& {Trieling}]{Bokhoven2009}
{\sc \au{{van Bokhoven}, L.~J.~A.}, \au{{Clercx}, H.~J.~H.}, \au{{van Heijst},
  G.~J.~F.} \& \au{{Trieling}, R.~R.}} \yr{2009}  \at{{Experiments on rapidly
  rotating turbulent flows}}.  \jt{Phys. Fluids}  \bvol{21}~(9),  \pg{096601}.

\bibitem[{Waleffe}(1993)]{Waleffe1993}
{\sc \au{{Waleffe}, F.}} \yr{1993}  \at{{Inertial transfers in the helical
  decomposition}}.  \jt{Phys. Fluids}  \bvol{5},  \pg{677--685}.

\bibitem[{Yarom} {\em et~al.\/}(2013){Yarom}, {Vardi} \& {Sharon}]{Yarom2013}
{\sc \au{{Yarom}, E.}, \au{{Vardi}, Y.} \& \au{{Sharon}, E.}} \yr{2013}
  \at{{Experimental quantification of inverse energy cascade in deep rotating
  turbulence}}.  \jt{Phys. Fluids}  \bvol{25}~(8),  \pg{085105}.

\bibitem[{Yeung} \& {Zhou}(1998)]{Yeung1998}
{\sc \au{{Yeung}, P.~K.} \& \au{{Zhou}, Y.}} \yr{1998}  \at{{Numerical study of
  rotating turbulence with external forcing}}.  \jt{Phys. Fluids}  \bvol{10},
  \pg{2895--2909}.

\bibitem[{Yoshimatsu} {\em et~al.\/}(2011){Yoshimatsu}, {Midorikawa} \&
  {Kaneda}]{Yoshimatsu2011}
{\sc \au{{Yoshimatsu}, K.}, \au{{Midorikawa}, M.} \& \au{{Kaneda}, Y.}}
  \yr{2011}  \at{{Columnar eddy formation in freely decaying homogeneous
  rotating turbulence}}.  \jt{J. Fluid Mech.}  \bvol{677},  \pg{154--178}.

\bibitem[Zel'dovich(1958)]{Zeldovich:1957zl}
{\sc \au{Zel'dovich, Ya.~B.}} \yr{1958}  \at{Electromagnetic interaction with
  parity violation}.  \jt{Sov. Phys. JETP}  \bvol{6},  \pg{1184}, [Zh. Eksp.
  Teor. Fiz. 33, 1531 (1957)].

\end{thebibliography}

\end{document}